\documentstyle[twoside,12pt,array,fleqn,epsf]{article}
\catcode `\@=11
\def\numberbysection{\@addtoreset{equation}{section}
         \renewcommand{\theequation}{\thesection.\arabic{equation}}}
\def\be{\begin{equation}}
\def\ee{\end{equation}}
\def\bd{\begin{displaymath}}
\def\ed{\end{displaymath}}
\newcommand{\ba}{\begin{eqnarray}}
\newcommand{\ea}{\end {eqnarray}}
\newcommand{\nn}{\nonumber}
\newcommand{\ket}[1]{|{#1}\rangle}

\def\a{\alpha}
\def\b{\beta}
\def\g{\gamma}

\def\d{\delta}

\def\sig{\sigma}
\def\eps{\epsilon}

\def\pr{\prime}
\def\ra{\rightarrow}

\def\half{\frac{1}{2}}

\numberbysection
\begin{document}
\pagestyle{plain}
\vspace* {10mm}
\begin{center}
\large
{\bf The free energy singularity of the asymmetric $6$--vertex model
and the excitations of the asymmetric $XXZ$ chain}
\end{center}
\begin{center}
\normalsize
       Giuseppe Albertini,
       Silvio Renato Dahmen and
       Birgit Wehefritz
        \\[1cm]
    {\it Universit\"{a}t Bonn,
                    Physikalisches Institut \\ Nu\ss allee 12,
                    D-53115 Bonn, Germany}\\[14mm]
{\bf Abstract}
\end{center}
\small
\noindent
%
%
We consider the asymmetric six--vertex model, {\it i.e.} the symmetric
six--vertex model in an external field with both horizontal and vertical
components, and the relevant asymmetric $XXZ$ chain. The model is widely
used to describe the equilibrium shape of a crystal. By means of the
Bethe Ansatz solution we determine the exact free energy singularity,
as function of both components of the field, at two special points on the
phase boundary. We confirm the exponent $\frac{3}{2}$ (already checked
experimentally), as the antiferroelectric ordered phase is reached 
from the incommensurate phase normally to the phase boundary, and we
determine a new singularity along the tangential direction. Both
singularities describe the rounding off of the crystal near a facet.
The hole excitations of the spin chain at this point on the phase
boundary show dispersion relations with the striking
form $\Delta E\sim (\Delta P)^{\half}$ at small
momenta, leading to a finite size scaling $\Delta E \sim N^{-\half}$
for the low--lying excited states, where $N$ is the size of the chain.
We conjecture that a Pokrovskii--Talapov phase transition is
replaced at this point by a transition with diverging correlation
length, but not classified in terms of conformal field theory.
%
%
\rule{5cm}{0.2mm}
\begin{flushleft}
\parbox[t]{3.5cm}{\bf PACS numbers:}
\parbox[t]{12.5cm}{05.70.Jk, 64.60.-i, 64.60.Fr, 75.10.Jm}
\end{flushleft}
\normalsize

\newpage
\pagestyle{plain}

\section{Introduction}

After the pioneering paper of Yang, Yang and Sutherland \cite{SYY} 
the asymmetric 
six-vertex model, i.e.\ the symmetric six-vertex model in a field, was
recently rediscovered because of its connection to a number of physically 
interesting problems, first among them the determination of the shape of
a crystal at equilibrium with its vapor phase \cite{van_Beijeren, Gwa_Spohn}.
This can be achieved by mapping the asymmetric $6$--vertex onto,
say, the $(001)$ facet of a bcc crystal under the condition that no
overhangs or voids are allowed (see e.g \cite{van_Beijeren} and
\cite{Jayaprakash_Saam} for details on the mapping). Excitations
in the vertex model correpond to small tilts away from the $(001)$
facet \cite{Jayaprakash_Saam}
and it can be shown \cite{Landau} that the 
free energy as function of the two components of the field
 gives exactly the equilibrium shape of the
crystal.

In its own, the asymmetric six-vertex model provides an interesting
2-dimensional system of interacting dipoles in an external field with
horizontal and vertical components ($h,v$). Fluctuating two--valued
variables (dipoles) are attached to the links of a two--dimensional
square lattice, and the model is defined by assigning a set of
Boltzmann weights (equivalently, interaction energies) to each
allowed vertex configuration (see fig. $1$). The transfer matrix can be 
diagonalized  exactly by the Bethe-Ansatz in its coordinate or algebraic 
version \cite{Lieb_Wu, Jayaprakash_Sinha}.
The phase diagram and the nature of the phase transitions
are well understood when $h=v=0$ (symmetric six-vertex), or when
$h=0$ and $v \neq 0$ \cite{Baxter, Lieb_Wu}. If $h,v \neq 0$, some
general features of the phase diagram have been described in \cite{SYY}
and the details of the calculation spelled out in \cite{Nolden}, but a few
questions have remained unanswered. It is known that, in the antiferroelectric
regime, with which we will be concerned in this paper, the free energy
$f(h,v)$ remains constant as  function of the field (`flat phase')
in a bounded region of the $(h,v)$ plane containing $h=v=0$.
This corresponds to the flat $(001)$ crystal plane. Beyond this
region, bounded by a curve $\Gamma$, the field is sufficiently strong 
to destroy the antiferroelectric order of the system, but not strong
enough to impose ferroelectric order, and an incommensurate phase
appears where the polarization (zero in the flat phase) changes
continuously with the field. 
Here the spectrum of the transfer matrix
is gapless with finite size corrections typical of the gaussian model
\cite{Noh_Kim}. The curve $\Gamma$ has been investigated only at a
few points and the nature of the phase transition was found to be of
a Pokrovskii--Talapov (PT) type \cite{PT}.
In this paper we determine the exact free energy singularity 
when approaching $\Gamma$ from the incommensurate phase,  which determines
the curvature of the crystal near the $(001)$ facet.
We
find two  special points $(h_c,v_c)$ (by symmetry under
arrow--reversal the same holds for  $(-h_c,-v_c)$, also on $\Gamma$),
where
\ba
f(h_c + \delta h, v_c) &=& f(h_c,v_c) - const (\d h)^{\frac{3}{2}}\nn\\
f(h_c,  v_c + \delta v) &=& f(h_c,v_c) - const |\d v|^3
\ea
These are points on $\Gamma$ where the tangent to $\Gamma$
is parallel to the $v-$axis. The exponent $3/2$ measures the rounding off
of the edges of the $(001)$ facet.

Even though our calculation has been carried out only at these  points 
of $\Gamma$,
the technique we present should work in general, and our result, which
generalizes Lieb and Wu's method and complements  the finite-size
techniques of Kim \cite{Noh_Kim}, strengthens the long held belief that
the exponent $3/2$ should govern the free energy singularity at every point
of the phase boundary \cite{van_Beijeren, Nolden}.
This exponent has been measured  \cite{Rottman}
in some  experiments with $Pb$ crystals some years ago. Our
results however show that along some tangential direction the exponent
$3$ should dominate, and this fact should be observable experimentally.

However, something more can be said about the nature of the phase transition
along $\Gamma$.

The method of mapping a 2d statistical system into a 1d quantum spin chain
has been fruitful and widely used in the past \cite{Kogut}.
We pursue it here, regardless of the fact that the relevant spin chain
which turns out to be the asymmetric $XXZ$ spin chain in a vertical field $V$,
is not hermitian \cite{McCoy_Wu}. We find that the flat phase corresponds
to a region in the $(h,V)$ plane where the ground state energy 
does not depend on the fields 
and where excitations  are massive. Along the transition line, analogue
of $\Gamma$, the excitations become massless, but the point $(h_c, V=0)$
(with its symmetric $(-h_c, V=0)$) is singled
out by the fact that dispersion
relations obey the striking law, at small momenta
\be
\Delta E \simeq (\Delta P)^{1/2}
\ee
and finite-size corrections for low-lying excitations scale like
\be
\Delta E \simeq N^{-1/2} 
\ee
if $N$ is the length of the chain.
At this point, analogue of $(h_c,v_c)$ for the statistical model, the vanishing
of the mass gap exhibits an exponent $1/2$ which does not appear
at any other point of the transition line in the $(h,V)$ plane. 
In section 5 we propose
an explanation of these results, arguing that at the point $(h_c,v_c)$
the transition occurs with a divergence of the correlation 
length in the correlator between two vertical arrows, while
everywhere else the transition is induced by level-crossing 
(Pokrovskii-Talapov) which
prevents the divergence of the same correlation length.

The paper is divided in 5 sections. In section 2 we give definitions
and summarize previously known results. In section 3 hole excitations
and the spectrum of the spin chain are studied and in section 4 the
method of Lieb and Wu is suitably extended to determine the free energy
singularity when both $h$ and $v$ are nonzero. Section 5 contains an
interpretation of the results.
%
%
\section{Definitions}

The model is a natural generalization of the well-known symmetric six-vertex 
model. Arrows are placed on the edges of an $N\times M$ square lattice
and Boltzmann weights
$R_{\alpha \alpha^{\prime}}^{\beta \beta^{\prime}}(u)$ are assigned to the 
vertices 
(see Fig.\ $1$) so that the row--to--row transfer matrices 
\be
T(u)_{\{\underline{\alpha}\},\{\underline{\alpha^{\prime}}\}}=
\sum_{\{\underline{\beta}\}}\prod_{k=1}^{N}
R_{\alpha_{k} \alpha_{k}^{\prime}}^{\beta_{k} \beta_{k+1}}(u)
\label{def1}
\ee
form a commuting family 
\bd
[T(u),T(u^{\prime})]=0
\ed
for any two values $u$, $u^\pr$ of the spectral parameter
\cite{Perk_Schultz}.  The associated (integrable) spin chain 
\be
{\cal H} = \sum_{j=1}^{N}\biggl[ \frac{\cosh \gamma}{2} 
(1+\sigma^z_j \sigma^z_{j+1})
              - e^{2h} \sigma^+_j \sigma^-_{j+1} -e^{-2h} 
\sigma^-_j \sigma^+_{j+1} \biggr] -V\sum_{j=1}^{N}\sigma_{j}^{z}
\label{def2}
\ee
is obtained from (\ref{def1}) by taking
the so-called extremely anisotropic limit ($u\rightarrow 0$) 
\bd
T(u)=e^{(v+h)\sum_{j=1}^{N}\sigma_{j}^{z}}\overline{T}(u) \;\;\;\;\;\;\;
{\cal H}= -V\frac{d}{dv}\log (e^{(v+h)\sum_{j=1}^{N}\sigma_{j}^{z}}) -
\sinh\gamma\frac{d}{du}
\log\overline{T}(u) \vert_{u=0}
\ed
$V$ breaks the $Z_2$ symmetry of spin reversal while $h$ breaks
parity invariance (see App. A for a complete discussion of
symmetries).

By means of the Bethe--Ansatz, eigenvalues of (\ref{def1}) and
(\ref{def2}) are found from the solution of a set of coupled equations
\be
\biggl[ \frac{\sinh (\frac{\gamma}{2} + \frac{i \alpha_k}{2})}
{\sinh(\frac{\gamma}{2} - 
\frac{i \alpha_k}{2})}\biggr]^N = (-1)^{n+1}e^{2hN} \prod_{l=1}^n 
\frac{\sinh(\gamma +
\frac{i}{2} (\alpha_k-\alpha_l))}{\sinh(\gamma -\frac{i}{2} 
(\alpha_k-\alpha_l))}\;\;\;\;\;\;k=1,2,\cdots,n
\label{def3}
\ee
and given respectively by
\ba
\Lambda (u)&=& e^{v (N-2n)}e^{hN}\biggl[
\frac{\sinh (\gamma - u)}{\sinh\gamma}\biggr]^{N}\prod_{j=1}^{n}
\frac{\sinh(\frac{\gamma}{2} + u - \frac{i\alpha_j}{2})}
{\sinh(\frac{\gamma}{2} - u + \frac{i\alpha_j}{2})}\nonumber\\
&&+e^{v (N-2n)}e^{-hN}\biggl[
\frac{\sinh u}{\sinh\gamma}\biggr]^{N}\prod_{j=1}^{n}
\frac{\sinh(\frac{-3\gamma}{2} + u - \frac{i\alpha_j}{2})}
{\sinh(\frac{\gamma}{2} - u + \frac{i\alpha_j}{2})} \label{def4} \nonumber\\
&=&\Lambda_{R}(u) + \Lambda_{L}(u)\\
E&=&N \cosh \gamma - \sum_{k=1}^{n}\frac{ 2 \sinh^2 \gamma}
{\cosh\gamma-\cos\alpha_k} -V(N-2n)\nonumber\\
&=& N\cosh \gamma + \sum_{k=1}^{n} e(\alpha_k)-V(N-2n)
\label{def5}
\ea
Here $n$ stands for the number of reversed spins (arrows) with respect to
the reference ferromagnetic state $\ket{\uparrow\uparrow\cdots\uparrow}$. It
is a conserved quantity since $S^{z}=\frac{1}{2}\sum_{j=1}^{N}\sigma_{j}^{z}$ 
commutes with ${\cal H}$ and $T(u)$.

Beside their energy, given by (\ref{def5}), the momentum can also be
computed. $\overline{T}^{-1}(0)$ yields the right--shift operator
$S=e^{-i{\mbox{\scriptsize P}}}$
\be
S=\ket{\a_1 ,\a_2 ,\cdots ,\a_{\mbox{\scriptsize M}}}=\ket{
\a_{\mbox{\scriptsize M}},\a_1 ,\a_2 ,\cdots ,\a_{\mbox{\scriptsize $M-1$}}}
\label{def6}
\ee
and from (\ref{def1}) and (\ref{def4}) one gets
\be
e^{-iP}=e^{-2hn}\prod_{j=1}^{n} \frac{\sinh(\frac{\gamma}{2} +
\frac{i\alpha_j}{2})} {\sinh(\frac{\gamma}{2} -\frac{i\alpha_j}{2})}
\;\;\;\;\;\;\;\;  P=-2inh -\sum_{j=1}^{n}p^{0}(\alpha_j )
\label{def7}
\ee
with
\bd
p^0(\alpha) =  -i \ln \biggl[\frac{\sinh (\frac{\gamma}{2} +
\frac{i \alpha}{2})} {\sinh(\frac{\gamma}{2} - \frac{i \alpha}{2})}
\biggr]
\ed

Unlike most integrable spin chains studied before, (\ref{def2}) is not
hermitian for $h\neq 0$, even though the statistical model has physically
sensible positive Boltzmann weights. The question arises whether (\ref{def1})
and (\ref{def2}) have a complete set of eigenvectors. By numerically 
diagonalizing the transfer matrix on small chains it appears that, even
though a few eigenvalues are degenerate in some charge sectors, the
eigenvectors are linearly independent. We will assume that  a complete
set of eigenvectors exists and it is given by the Bethe Ansatz.

We first summarize, for the sake of completeness, some already known facts
that have been published elsewhere, beginning with
the spin chain because, although the Bethe Ansatz equations are
identical, the form of the energy contribution from each rapidity
$\alpha_k$ is simpler.
Taking the logarithm of (\ref{def3}) in the usual way we get
\be
p^0(\alpha_k)-\frac{1}{N} \sum_{l=1}^n \Theta(\alpha_k-\alpha_l) + 2ih =
\frac{2 \pi}{N} I_k\;\;\;\;\;\;\;\;\;  k=1,2,\cdots ,n
\label{def8}
\ee
where $I_k$ is half-odd (integer) if $n$ is even (odd) and
\bd
\Theta(\alpha)  =  -i \ln \biggl[\frac{\sinh(\gamma +
\frac{i\alpha}{2})}{\sinh(\gamma -\frac{i\alpha}{2})}\biggr]
\ed
We define $p^0(0)=\Theta (0) = 0$ and cuts are chosen to run from $i\gamma$ to
$i\infty$ and from $-i\gamma$ to $-i\infty$ for $p^0(\alpha)$; from $2i\gamma$
to $i\infty$ and from $-2i\gamma$ to $-i\infty$ for $\Theta(\alpha)$. Notice
from (\ref{def3}) that $Re(\alpha)\in [-\pi,\pi]$ so $Re(\alpha_k - \alpha_{l})
\in [-2\pi,2\pi]$. The cuts are chosen so that $\Theta(\alpha)$ is analytical
in $-2\gamma < Im(\alpha) < 2\gamma$ and real, monotonically increasing when
$\alpha \in [-2\pi,2\pi]$. With these conventions, the ground state at $h=0$
in each sector of fixed $S^{z}=\frac{N}{2}-n$ corresponds to a
sequence of $n$ consecutive numbers $\{ I_{k}\}$ in (\ref{def8}), from
$-\frac{n-1}{2}$ to $\frac{n-1}{2}$ symmetric around $0$ \cite{YY}.
The rapidities $\{ \alpha_j \}$ are real and distributed symmetrically around
$\alpha = 0$ too. As $h\neq 0$, the rapidities move into the complex
plane along a curve $C$. In a standard way \cite{Nolden}, one gets in the
thermodynamic limit from (\ref{def8}) 
\be
p^0(\alpha)-\frac{1}{2 \pi } \int_C d\beta \Theta(\alpha -\beta) 
R(\beta) + 2ih=2\pi x \;\;\;\;\;\;\;
-\frac{1-y}{4}\leq  x \leq \frac{1-y}{4}\\
\label{def9}
\ee
$x$ is the real parameter of the curve, $y$ is the polarization
defined through
\be
y= \lim_{N \rightarrow \infty} \frac{2S^{z}}{N} = \lim_{N \rightarrow \infty}
(1-\frac{2n}{N})
\label{def10}
\ee
and the rapidity density $R(\alpha_l ) =
\lim_{N\rightarrow \infty}\frac{2\pi}{N(\alpha_{l+1}-\alpha_l )}$
is determined by solving the integral equation
\be
\xi(\alpha )-\frac{1}{2 \pi } \int_C d\beta K(\alpha -\beta)R(\beta)=
R(\alpha)\\
\label{def11}
\ee
where
\bd
\xi(\alpha ) = \frac{dp^{0}(\alpha)}{d\alpha}; \;\;\;\;\;\;\;\;\;
K(\alpha ) = \frac{d\Theta (\alpha )}{d\alpha}
\ed
The energy and the polarization are thus given by
\ba
\lim_{N\rightarrow \infty} \frac{E}{N}&=&\cosh \gamma +\frac{1}{2\pi}
\int_C d\alpha e(\alpha) R(\alpha) - Vy\label{def12}\\
&&\nonumber\\
\frac{1-y}{2}&=&\frac{1}{2\pi}\int_C d\alpha R(\alpha)
\label{def13}
\ea 
Some preliminary information about the shape of $C$ and its location in the
complex plane can be obtained by solving (\ref{def3}) numerically. We
take as initial solution that
composed of real roots and corresponding to the ground state, at
fixed $S^{z}$, for $h=0$. By the Perron--Frobenius theorem
\cite{YY, Gantmacher}, the relevant eigenstate
remains the ground--state at fixed $S^{z}$ even when $h\neq 0$, and it is real.
It appears that for $h>0$ ($h<0$) the rapidities move into the lower (upper)
half-plane as shown in fig. $2$.

The curve $C$ is invariant under $\alpha\rightarrow -\alpha^{\ast}$, which is to
be expected, being this transformation a symmetry of (\ref{def3}), so we will
set $A=-a+ib$ and $B=a+ib$ to be the endpoints of the curve. Note that this
property makes $E$ real, as it should.

Strictly speaking $R(\alpha)$ is defined on $C$ only,
but (\ref{def11}) can be used to define it outside of $C$. If $C$ is
contained in the strip $-\gamma< Im(\alpha) < \gamma$, $R(\a)$ is
analytic in $-\gamma < Im (\alpha) < \gamma$, but it inherits the poles
of $\xi (\a)$ at $\pm i\gamma$.
Let us consider the curve for which $a=\pi$. In this case, 
since $R(\alpha )$ is $2\pi$--periodic, (\ref{def11}) can be solved
straightforwardly by Fourier transform. The solution is
\be
R(\alpha )=\sum_{n}\frac{e^{-in\alpha}}{2\cosh \gamma n}
\;\;\;\;\;\;\;\;\;\;\;\; -\gamma < Im(\alpha )<\gamma
\label{def14}
\ee
Here and in the following, sums are understood to run from $-\infty$
to $\infty$ unless otherwise stated. Beyond this strip (\ref{def14}) can be
expressed using elliptic
functions. Introducing the complete elliptic integrals of the first kind
$I$ ($I'$) of modulus $k$ ($k'$), with $k^2 + k'^{2} = 1$ \cite{Erdelyi},
related to $\gamma$ by
\bd
\frac{I'(k)}{I(k)} = \frac{\gamma}{\pi}
\ed
the solution of (\ref{def11}) in a wider domain reads \cite{Gaudin, Erdelyi}
\be
R(\alpha ) = \frac{I(k)}{\pi}{\mbox{\sf dn}}
\biggl(\frac{I(k)\alpha}{\pi}; k\biggr)
\label{def15}
\ee
Notice the presence of a pole at $\alpha = \pm i\gamma$, inherited from
$\xi (\alpha)$, which prevents the convergence of
(\ref{def14}) beyond the smaller domain. The energy
remains constant at its value for $h=0$
\bd
e_0 = \lim_{N\rightarrow \infty} \frac{E_0}{N} = \cosh\gamma-
2\sinh\gamma\sum_{n}\frac{e^{-\gamma |n|}}{2\cosh \gamma n}
\ed
and from (\ref{def13}) $y=0$. In fact the solution considered here has
$n = \frac{N}{2}$ rapidities $(S^{z}=0)$ and, as it will be shown in the 
next section,
$E_0$ is the ground state energy for $h$ and $V$
sufficiently close to $0$.

As to the precise position of the curve, one
has to revert to (\ref{def9}). Since
$\Theta(\alpha +2\pi ) =\Theta(\alpha )+2\pi$, we use the expansion
\bd
\Theta (\alpha ) = \alpha +i\sum_{n\neq 0}\frac{e^{-in\alpha -2\gamma |n|}}{n}
\;\;\;\;\;\;\;\;\;\;\;\; -2\gamma < Im(\alpha) <2\gamma
\label{thetaseries}
\ed
and
\be
p^{0}(\alpha ) = \alpha +i\sum_{n\neq 0}\frac{e^{-in\alpha -\gamma |n|}}{n}
\label{def16}
\ee
and we introduce
\be
p(\alpha ) = \frac{\alpha}{2}+i\sum_{n\neq 0}\frac{e^{-in\alpha}}
{2n\cosh\gamma n} = {\mbox{\sf am}}\biggl(\frac{I(k)\alpha}{\pi}; k\biggr)
\label{def17}
\ee
The series in (\ref{def16}) and (\ref{def17}) are
certainly convergent when $-\gamma < Im(\alpha) <\gamma$ but they converge also
at $\alpha = \pm\pi\pm i\gamma$, because of the alternating sign. Eq.
(\ref{def9}) reduces to
\bd
p(\alpha ) + \frac{ib}{2} + i\sum_{n\neq 0}(-)^{n}\frac{e^{nb}}
{2n\cosh\gamma n} + 2ih = 2\pi x \;\;\;\;\;\;\;\;\;\;\;
-\frac{1}{4}\leq x \leq \frac{1}{4}
\ed
Specialization to the endpoints permits relating the value of $h$ to $b$
\cite{Lieb_Wu, Nolden}
\be
h(b) = -\frac{b}{2}-\sum_{n=1}^{\infty}(-)^{n}\frac{\sinh nb}{n\cosh n\gamma}
\label{def18}
\ee
so that the final equation of the curve is
\bd
p(\alpha )+ih = 2\pi x \;\;\;\;\;\;\;\;\;\;\;
-\frac{1}{4} \leq x \leq \frac{1}{4}
\label{curve}
\ed
Notice that the points on the curve are characterized by
\be
Im(p(\a )) +h = 0
\label{def19}
\ee
We set $h_c = h(b=-\gamma )$. One
might suspect that, when $h>h_c$, the endpoints would remain at
$a=\pi$ but with $b<-\gamma$ (or $b>\gamma$ if $h<-h_{c}$). If $C$
does not cross the point $-i\gamma$ where $\xi(\a)$ has a pole, it can
always be deformed to the real axis in (\ref{def11})
\be
R(\a) + \frac{1}{2\pi}\int_{-\pi}^{\pi}duK(\a -u)R(u) = \xi (\a)
\label{def20}
\ee
so that the solution is still given by $\frac{I(k)}{\pi}{\mbox{\sf dn}}
\biggl(\frac{I(k)}{\pi}\a ;k\biggr)$, but the expansion (\ref{def14}) is no
longer useful. To find the $h(b)$ relation, we close $C$ in (\ref{def9}) to
the real axis, and take $\a =A$
\be
p^0(A)-\frac{1}{2\pi} \int_{-\pi}^{\pi} du\Theta(A-u)R(u)
+\int_{-\pi}^{A} d\beta R(\beta ) + 2ih =-\frac{\pi}{2}
\label{def21}
\ee
The integral of $R(\a)$ is obtained by integrating both sides of
(\ref{def20}), and from (\ref{def21}) we conclude
\be
2h(b) = b + 2\ln\frac{\cosh\biggl(\frac{\gamma}{2}-\frac{b}{2}\biggr)}
{\cosh\biggl(\frac{\gamma}{2}+\frac{b}{2}\biggr)} + 2\sum_{n>0}
(-)^n\frac{e^{-2n\gamma}\sinh nb}{n\cosh n\gamma}
\label{def22}
\ee
which reduces to (\ref{def18}) when $|b|\leq \gamma$. Eq. (\ref{def22})
though cannot give the right dependence $h(b)$ at $|b|>\gamma$, because $h(b)$
decreases when $b<-\gamma$ and increases for $b>\gamma$, going back
to the range of values it had as $b\in [-\gamma,\gamma]$. Clearly
the initial assumption $a=\pi$ cannot be correct. To gain more insight
we resort as usual to the numerical solution of (\ref{def3}) which shows
that, as $h>h_c$, the curve with $y=0$ has endpoints at (see table $2$. Note
that all extrapolations presented in the tables have been done using data
up to $80$ sites. We give however only the first values, up to $N=40$,
for they already show clearly that the values are converging towards a
limit. All tables have been calculated for $\cosh\gamma = 21$, where
$\exp (2h_c)= 10.51787$).
\bd
b=-\gamma \;\;\;\;\;\;\;\;\; a<\pi
\ed
This (new) result will be used in the calculation of the free energy
singularity in Section 4.

Turning next to the statistical model, the largest eigenvalue of the 
transfer matrix yields the free energy
per site (we drop here the inessential factor $\b$)
\bd
f(u,\gamma,h,v)=-\lim_{N\ra\infty}\frac{\ln \Lambda_{0}(u,\gamma,h,v)}{N}
\ed
whose value, as $N\ra\infty$, is dominated by the largest of the two
limits
\begin{eqnarray*}
\lim_{N\ra\infty}\frac{1}{N}\Lambda_{\mbox{\scriptsize R}}(u)=
F_{\mbox{\scriptsize R}}(u,\gamma,h,y) +vy &=& h
+\ln\frac{\sinh (\gamma -u)}
{\sinh\gamma} +vy\nonumber\\
&&+\frac{1}{2\pi}\int_C d\a R(\a )f_{\mbox{\scriptsize R}} (\a ;u)\\
\lim_{N\ra\infty}\frac{1}{N}\Lambda_{\mbox{\scriptsize L}}(u)=
F_{\mbox{\scriptsize L}}(u,\gamma,h,y) +vy &=& -h
+\ln\frac{\sinh (u)}
{\sinh\gamma} +vy\nonumber\\
&&+\frac{1}{2\pi}\int_C d\a R(\a )f_{\mbox{\scriptsize L}} (\a ;u)
\end{eqnarray*}
where we have defined
\begin{eqnarray*}
f_{\mbox{\scriptsize R}} (\a ;u)&=&\ln \frac{\sinh(\frac{\gamma}{2}
+ u -\frac{
i\alpha_j}{2})} {\sinh(\frac{\gamma}{2} - u + \frac{i\alpha_j}{2})}\\
f_{\mbox{\scriptsize L}} (\a ;u)&=&\ln
\frac{\sinh(\frac{-3\gamma}{2} + u - \frac{i\alpha_j}{2})}
{\sinh(\frac{\gamma}{2} - u + \frac{i\alpha_j}{2})}
\end{eqnarray*}
If $F_{\mbox{\scriptsize R}}$, $F_{\mbox{\scriptsize L}}$ are known, and
we call $F$ the dominant one, the equilibrium value of $y$ and the free
energy are determined by the minimum condition
\be
f(u,\gamma,h,v) = \min_{-1\leq y\leq 1}
\{-F(u,\gamma,h,y)-vy\}
\label{def23}
\ee
When $-h_c\leq h\leq h_c$ and for small enough values of $v$, the
state defined by $C$ ($a=\pi ; -\gamma \leq b\leq \gamma$) also yields
the largest eigenvalue of the transfer matrix. The free energy
\footnote{The only part in which our analysis differs from
\cite{Nolden} is that $\Lambda_{\mbox{\scriptsize R}}$ is exponentially larger
for $d<\gamma -2u$ and $\Lambda_{\mbox{\scriptsize L}}$ for  $d>\gamma -2u$
where $id$ is the point in which $C$ crosses the imaginary axis. A comparison
of (\ref{def24}) with \cite{Nolden} should take into account the
different normalization of the Boltzmann weights.}
\be
f(u,\gamma,h,v) = -2\sum_{n=1}^{\infty}\frac{e^{-2\gamma n}}{n\cosh \gamma n}
\sinh (nu)\sinh n(\gamma -u)
\label{def24}
\ee
is constant in a whole region of the $(h,v)$ plane bounded by a curve
$\Gamma$ ('flat' phase). The parametric equation $(h(b),v(b))$ of
$\Gamma$ is given by (\ref{def20}) and by
\bd
v(b) = -\frac{\partial F}{\partial y}\bigg
\vert_{\mbox{\scriptsize h fixed, y=0}}
\ed
which can be explicitly computed
\be
2v(b) = \gamma -|\gamma -2u +b | + 2\sum_{n=1}^{\infty}\frac{(-)^n}{n}
\frac{\sinh [n(\gamma-|\gamma-2u-b|)]}{\cosh n\gamma}\;\;\;\;\;\;\;
-\gamma\leq b\leq\gamma
\label{def25}
\ee
The other half of the curve $\Gamma$ (see fig. $3$) can be recovered
from the symmetry $f(-h,-v)=f(h,v)$ (see Appendix A).
Most of these results have been obtained elsewhere and we have presented
them here only for the sake of completeness. It should be pointed out that
the fact that the ground state energy does not depend
on $h$ is simply a consequence of the analogous property of the free energy.
%
%
\section{Hole excitations of the spin chain}

To understand the nature of the phase transition along $\Gamma$ we
turn to the calculation of the excitation energies, and we set $V=0$,
since the role of $V$ is simply to shift the spectra at $S^{z}\neq 0$.
As proven in appendix A it is sufficient to consider $h\geq 0$.

A complete treatment of the spectrum should rely on the classification of
all possible solutions of (\ref{def3}). This is usually done in the
framework of the string hypothesis, according to which complex rapidities
(at $h=0$) have an imaginary part which tends to well defined values in
the thermodynamic limit. Exceptional solutions other than
strings, but that still appear in complex conjugate pairs, can be handled
in a similar way \cite{Babelon_deVega_Viallet}. Yet, from the numerical
analysis of (\ref{def3}), it appears that strings do not survive at
moderately strong
values of $h$. Therefore we shall limit our calculation to the so--called
hole excitations, that is holes in the ground state distribution of rapidities,
occurring in sectors with $S^{z}>0$ ($n<n_0 = \frac{N}{2}$). We introduce
the counting function
\be
Z(\alpha , \{\alpha_j\} ) = \frac{p^{0}(\alpha )}{2\pi}-\frac{1}
{2\pi N}\sum_{j=1}^{n}\Theta (\alpha -\alpha_j ) + i\frac{h}{\pi}
\label{def26}
\ee
so that (\ref{def8}) is rewritten
\bd
Z(\alpha_k ) = \frac{I_k}{N}
\ed
Set $vac$ $=$ number of vacancies available for the quantum numbers
$\{I_k\}$. With the usual hypothesis that $Z(\alpha )$ be monotonically
increasing, we have
\be
\Delta Z\stackrel{\mbox{\scriptsize\sf def}}{=}
Z(\alpha )\vert_{\mbox{\scriptsize Re($\alpha$)$=\pi$}} -
Z(\alpha )\vert_{\mbox{\scriptsize Re($\alpha$)$=-\pi$}} = \frac{vac}{N}
\label{def27}
\ee
On the other hand, from (\ref{def26}), we find $\Delta Z = 1-\frac{n}{N}$. For
the ground state $n=n_0 = \frac{N}{2}$ and so $vac=n_0$, {\it i.e.} the
available vacancies are all filled. For $n$
rapidities, $n=n_0 -r$ where $r=1,2,\cdots$ we have
\bd
\Delta Z = 1-\frac{n_0 -r}{N}\;\;\;\;\;\;\;\;\ vac = n_0 + r
\ed
so that $n_0 + r$ vacancies are partially filled with ($n_0 - r$) $I_k$'s,
leaving $N_h = 2r$ holes.

We will resort to the 'backflow method' \cite{Gaudin}
\cite{Bogoliubov_Izergin_Korepin} in dealing
with (\ref{def5}), (\ref{def26}) and (\ref{def7}) in the limit $N\ra\infty$.
The calculation differs slightly for the $2$ cases $r=$ even or odd, but the
results are identical and we will present the case $r=$ even only. If this
is the case, then $n=n_0$ (mod$2$) and the quantum numbers $\{I_k \}$ of the
excited state have the same oddness of the quantum numbers $\{I_{k}^{0}\}$ of
the ground state. We assume that the $r$ additional vacancies for $\{I_k \}$
are placed $\frac{r}{2}$ to the left and $\frac{r}{2}$ to the right of the
sequence (see fig. $4$).

We call $\{\b_{j}^{(1)}\}$ the $\frac{r}{2}$ additional rapidities at the
left edge, $\{\b_{j}^{(2)}\}$ the $\frac{r}{2}$ additional rapidities
at the right edge and $\{\a_{j}^{(h)}\}$ the $N_h$ holes.
Then, for the ground state
\ba
Z_{0}(\alpha )&=&\frac{p^{0}(\alpha )}{2\pi}-\frac{1}{2\pi N}
\sum_{j=1}^{n_0}\Theta (\alpha -\alpha_{j}^{0})+i\frac{h}{\pi}
\\
\label{def28}
Z_{0}(\alpha_{k}^{0})&=&\frac{I_{k}^{0}}{N}
\label{def29}
\ea
and for the excited state, adding and subtracting the holes
\ba
Z(\alpha )&=&\frac{p^{0}(\alpha )}{2\pi}-\frac{1}{2\pi N}
\sum_{j=1}^{n_0}\Theta (\alpha -\alpha_{j})+
\frac{1}{2\pi N}\sum_{j=1}^{N_h}\Theta (\alpha -\alpha_{j}^{(h)})
\nonumber\\
&-&\frac{1}{2\pi N}\sum_{j=1}^{\frac{r}{2}}\biggl[
\Theta (\alpha -\b_{j}^{(1)})+\Theta (\alpha -\b_{j}^{(2)})\biggr]
+i\frac{h}{\pi}\\
\label{def30}
Z(\alpha_{k})&=&\frac{I_{k}}{N}\;\;\;\;\;\;\;\;
Z(\alpha_{j}^{(h)})= \frac{I_{j}^{(h)}}{N}
\label{def31}
\ea
As $N\ra\infty$  $\{\b_{j}^{(1)}\} \ra A$ and $\{\b_{j}^{(2)}\}\ra B$.
Subtracting (\ref{def28}) from (\ref{def30}) and retaining terms of
order $\frac{1}{N}$ (this is a standard Bethe Ansatz calculation)
\cite{Bogoliubov_Izergin_Korepin} we find that
\bd
j(\alpha_l ) = \lim_{N\ra\infty}\frac{\alpha_l -\alpha_{l}^{0}}
{\alpha_{l+1}^{0} -\alpha_{l}^{0}}
\ed
satisfies
\ba
j(\alpha ) +  \frac{1}{2 \pi }\int_C d\beta K(\alpha -\beta)j(\beta )&=&
-\frac{1}{2\pi}\sum_{j=1}^{N_h}\Theta (\alpha -\alpha_{j}^{(h)})\nonumber\\
&+&\frac{r}{4\pi}\biggl[
\Theta (\alpha -B)+\Theta (\alpha -A)\biggr]
\label{def32}
\ea
with
\ba
\Delta E &=& \int_C d\alpha e^{\pr}(\alpha )j(\alpha )-
\sum_{j=1}^{N_h}e(\alpha_{j}^{(h)}) +\frac{r}{2}
\biggl( e(A)+e(B)\biggr)\nonumber\\
\Delta P &=& ihN_{h} - \int_C d\alpha \xi (\alpha)j(\alpha )
+\sum_{j=1}^{N_h}p^{0}(\a_{j}^{(h)}) -\frac{r}{2}\biggl(
p^{0}(A)+p^{0}(B)\biggr)\nonumber
\ea
Eq. (\ref{def32}) defines the analytical properties of $j(\alpha )$ in the
complex plane. Since for $0\leq h\leq h_c$ the curve $C$ is contained in
$-\gamma\leq Im(\a )\leq 0$, $j(\a )$ is certainly analytic in
$-2\gamma <Im(\a )<\gamma$ and the curve can be closed to the real axis.
Noticing, from (\ref{def32}), that $j(\a +2\pi)-j(\a )=-\frac{N_h}{2}$ we get,
with $u\in [-\pi ,\pi]$
\ba
j(u)&+&\frac{1}{2\pi}\int_{-\pi}^{\pi}dvK(u-v)j(v)=\frac{N_h}{4\pi}
\Theta (u+\pi)-\frac{1}{2\pi}\sum_{j=1}^{N_h}\Theta (u-\a_{j}^{(h)})
-\frac{r}{2}\label{def33} \\
\Delta E &=& \int_{-\pi}^{\pi} du e^{\pr}(u)j(u)+
\frac{N_h}{2}e(\pi )-\sum_{j=1}^{N_h}e(\a_{j}^{(h)}) \label{def34}\\
\Delta P &=& ihN_{h}- \int_{-\pi}^{\pi} du\xi (u)j(u)
+\sum_{j=1}^{N_h}p^{0}(\a_{j}^{(h)})
\label{def35}
\ea
Equation (\ref{def33}) can be solved by Fourier transform, paying
attention to the fact that $j(u)$ is not periodic, but obeys the 
quasiperiodicity condition $j(u+2\pi )-j(u)=-\frac{N_h}{2}$. Alternatively,
and with identical results, the symmetric integral operator
$({\bf 1}+\frac{1}{2\pi}{\bf K})$ at the left side of (\ref{def33}) can be
formally inverted and the solution plugged in (\ref{def34}) and (\ref{def35})
\cite{Bogoliubov_Izergin_Korepin}. The result has the usual additive form
\be
\Delta E = 2\sinh\gamma\sum_{j=1}^{N_h}\eps (\a_{j}^{(h)})
\label{def36}
\ee
where the dressed energy $\eps (u)$ satisfies
\bd
\eps (u) + \frac{1}{2\pi }\int_{\pi}^{\pi} dv K(u-v)\eps (v) =\xi (u)
\ed
and therefore coincides with $R(u)=\frac{I(k)}{\pi}{\mbox{\sf dn}}\biggl(
\frac{I(k)u}{\pi};k\biggr)$. As to the momentum, one has
\be
\Delta P = \sum_{j=1}^{N_h}\biggl[p(\a_{j}^{(h)})+ih
\biggr]
\label{def37}
\ee
where $p(\a )$ has been defined in (\ref{def17}). The calculation for $r$ odd
differs slightly in the intermediate steps but also yields (\ref{def36}) and
(\ref{def37}), which are then true for $N_h$ arbitrary (but obviously even).
The fact that $N_h$ is even was missed in \cite{wir} where, following
the same assumption made in \cite{Gaudin}, one hole was kept fixed at
the edge. In other words, only a subset of the $2$--hole band of states
was dealt with \footnote{One of the authors (GA) is grateful to prof.
C. Destri for pointing this out.}.

Eq. (\ref{def36}) and (\ref{def37}) are simple generalizations of their limit
at $h=0$, since $h$ appears only additively in $\Delta P$ and, implicitly,
in the position of the hole which is bound to be on the curve $C$. This
simple dependence could not be derived immediately from (\ref{def5})
and (\ref{def7}) because the
rapidities $\{\a \}$ depend on $h$ in a nontrivial way through (\ref{def3}),
and only the explicit calculation guarantees that (\ref{def36}) and
(\ref{def37}) are correct.

Several comments are in order. Unlike the energy, which being the eigenvalue
of a non--hermitian operator can, and indeed does have an imaginary part, the
momentum must be real. This is guaranteed by (\ref{def19}),
since $\a^{(h)}\in C$.
It is well known that, from (\ref{def7}), (\ref{def8}) and the oddness of
$\Theta (\a )$, the momentum can be obtained by summing (\ref{def8}) over $k$
\be
P = -\frac{2\pi}{N}\sum_{k=1}^{n}I_k
\label{def38}
\ee
The momentum of the ground state $P^{0}$ is therefore always zero, while
the momentum of an excited state is
\bd
P= \frac{2\pi}{N}\sum_{k=1}^{N_h}I_{k}^{(h)}
\ed
from which one sees that, as $N\ra\infty$, $-\frac{\pi}{2}\leq\Delta P
({\mbox{\sf hole}})\leq\frac{\pi}{2}$. This is also confirmed by (\ref{def17})
and (\ref{def37}). The dispersion relations are obtained by eliminating
$\a^{(h)}$ in (\ref{def36}) and (\ref{def37}) using
\bd
{\mbox{\sf dn}}(\a ;k) = \sqrt{1-k^{2}\sin^{2}({\mbox{\sf am}}(\a ;k)}
\ed
which yields
\be
\Delta E (\Delta P) = m_0 \sqrt{1-k^{2}\sin^{2}\biggl(\Delta P -ih\biggr)}
\;\;\;\;\;\;\;\;\;   m_0 = 2\sinh\gamma\frac{I(k)}{\pi}
\label{def39}
\ee
An apparent discrepancy with Gaudin's result (for the case $h=0$)
is clarified in appendix B.
The ${\mbox{\sf dn}}(\a )$ function 
(and consequently $\eps (\a )$) has a non negative
real part in the rectangle $[-\pi ;-\pi -i\gamma ; \pi -i\gamma ; \pi ]$
and this lifts the ambiguity in the sign of (\ref{def39}). 
It also confirms that
the choice of the ground state was correct, because the real part of the
energy, at least under 'small' variations (a countable number of holes),
increases. The mininum of
$Re(\Delta E)$ is reached when $\a^{(h)}=A$ or $B$. When this happens
$\Delta P =\pm\frac{\pi}{2}$ and the gap in the spectrum is (remember that
$2$ holes are present in the lowest excited state)
\be
\Delta E ({\mbox{\sf gap}}) = 2m_0 \sqrt{1-k^{2}\cosh^{2}(h)}
\label{def40}
\ee
which guarantees that the 'mass gap' is real (see table $3$ for a comparison
with numerical results). In particular it vanishes
at $b=-\gamma$, that is when $A=-\pi-i\gamma$, $B=\pi-i\gamma$ and
\bd
h = h_c = \frac{\gamma}{2} + \sum_{n\neq 0}(-)^{n}
\frac{\sinh (n\gamma )}{n\cosh (n\gamma )}
\ed
This is most easily seen from (\ref{def36})
\bd
\Delta E ({\mbox{\sf gap}}) = 2m_0 {\mbox{\sf dn}}\biggl(
\frac{I(k)}{\pi}(\pm\pi -i\gamma); k\biggr) = 2m_0 {\mbox{\sf dn}}
(\pm I(k)-i I^{\pr}(k); k) = 0
\ed
Therefore, from (\ref{def40}) an alternative equation for $h_c$ is
\bd
\cosh (h_c) = \frac{1}{k}
\ed
It is particularly interesting to see how the mass gap vanishes as
$h\ra h_c$
\be
\Delta E ({\mbox{\sf gap}}) \sim 2^{\frac{3}{2}}m_0\sqrt{k^{\pr}}
(h_c -h)^ \half + O(h_c -h)
\label{def41}
\ee
The vanishing with an exponent $\half$ is peculiar of the point under
consideration, as it will appear clear from the general case, to be
discussed later, which includes the vertical field $V$. Finally, we
specialize (\ref{def40}) at $h=h_c$. Then the hole excitations are massless
and if we set $\Delta P = -\frac{\pi}{2} +\eps$ or $\Delta P = \frac{\pi}{2}
-\eps$, $0<\eps<<1$, we get, respectively  
\be
\Delta E (\Delta P)\sim  2m_0\sqrt{\mp 2ik^{\pr}}\;\;\eps ^{\half}
\label{def42}
\ee
This dispersion relation is certainly surprising and reflects itself
in the peculiar behavior of the finite size corrections of the low--lying
energy gaps at $h=h_c$. In marked contrast with the ${\cal O}(\frac{1}{N})$
scaling typical
of spin chains which describe conformally invariant models in the continuum
limit \cite{Cardy}, we find
\bd
\Delta E \sim \frac{c}{N^{\half}} + O(N^{-1})
\ed
where $c$ depends on the state under consideration. The momentum being
quantized in units of $\frac{2\pi}{N}$ on the finite lattice (\ref{def38}),
this behavior is well in agreement with (\ref{def42}).

The sector $S^{z}=0$ deserves a special comment. Not knowing what takes
the place of strings, the analysis of the excitations has been necessarily
numerical. Two things have been determined. Setting $E_{0}(S^{z}=0,N,h)$
and $E_{1}(S^{z}=0,N,h)$ to be respectively the ground state and the
lowest--lying of the first band of excited states
in the sector $S^{z}=0$, on a chain of $N$ sites, and at
fixed horizontal field $h\leq h_c$, we found
\bd
\Delta E (S^{z}=0,h) =\lim_{N\ra\infty}\biggl[E_{1}(S^{z}=0,N,h) -
E_{0}(S^{z}=0,N,h)\biggr]
\ed
to be positive, non--zero for $h<h_c$ and
\bd
\lim_{h\ra h_{c}^{-}}\Delta E (S^{z}=0,h) = 0
\ed
so that, even in this sector, the spectrum becomes massless at $h=h_c$ (see
table $4$).
Secondly, the ${\cal O}(N^{-\half})$ scaling is preserved at $h_c$
\bd
E_{1}(S^{z}=0,N,h_c ) - E_{0}(S^{z}=0,N,h_c )\sim\frac{c}{N^{\half}}
\;\;\;\;\;\;\; N \gg 1
\ed
$E_1$ must not be confused with the other (degenerate in the
thermodynamic limit) ground state that appears in this sector at
momentum $P=\pi$ and is responsible for the spontaneous breaking
of the arrow--reversal symmetry in the symmetric $6$--vertex model
\cite{Baxter, Gaudin}. See table $5$ for a comparison with numerical
results.
Since we do not study the order parameter (staggered polarization)
this state will not be discussed here.

We can now reintroduce $V$, whose effect is to shift the spectra at
$S^{z}\neq 0$. The mass gap for a state
with $n=n_{0}\pm r$, and consequently $2r$ holes
is easily read from (\ref{def5}) and (\ref{def40}) 
\be
\Delta E ({\mbox{\sf gap}};n) = 2rm_0 \sqrt{1-k^{2}\cosh^{2}(h)} -2V(\pm r)
\label{def43}
\ee
An alternative way to reach the boundary with the massless phase is to have
a sufficiently large $|V|$. From (\ref{def43}) the crossing occurs at
\be
V = \pm m_0 \sqrt{1-k^{2}\cosh^{2}(h)}
\label{def44}
\ee
and moves the ground state to sectors of $S^{z} > 0$ ({\it i.e.} $n<n_0$) if
$V>0$, and to sectors of $S^{z} < 0$ ({\it i.e.} $n>n_0$) if $V<0$, as it was
intuitively predictable from (\ref{def5}). Notice that, unlike what happens in
(\ref{def41}), the mass gap goes to zero linearly in $V$ or linearly in $h$ if
$V\neq 0$ were kept fixed and $h\ra h(V)$, where $ h(V)$ is defined by
(\ref{def44}). The point $V=0$, $h=h_c$ (or equivalently $h=-h_{c}$),
where the exponent $\half$ of (\ref{def41}) appears, is clearly special.
Even if it were approached by changing $h$ and $V$ simultaneously, the term
$(h_c -h)^{\half}$ would dominate over the linear term in $V$.
There is no way to erase this effect because it is impossible to reach
$(h_c ,V=0)$ by changing $V$ only: the line $h=h_c$ in the $(h,V)$ plane
is tangent to the phase boundary curve defined by (\ref{def44}).

This result may look odd, because the energy difference between sectors
of different $S^{z}$ corresponds to the step free energy for the statistical
model, and from (\ref{def4}) it is hard to see how it could vanish other than
linearly. Yet it is readily seen that the phenomenon is not an artifact
of the spin chain. An explicit calculation of the step free energy
\bd
f_{\mbox{\scriptsize step}} = -\biggl[ \ln\Lambda_{\mbox{\scriptsize max}}
(S^{z}=1)- \ln\Lambda_{\mbox{\scriptsize max}}(S^{z}=0)\biggr]
\ed
can be bypassed observing that the vanishing of $f_{\mbox{\scriptsize step}}$
signals the transition to the incommensurate phase and therefore must be
given by (\ref{def25})
\ba
f_{\mbox{\scriptsize step}}&=&2v -\gamma +
|\gamma -2u-b|\nonumber\\
&& - 2\sum_{n=1}^{\infty}\frac{(-)^n}{n}
\frac{\sinh [n(\gamma-|\gamma-2u-b|)]}{\cosh n\gamma}\;\;\;\;\;\;\;
-\gamma\leq b\leq\gamma
\nn
\ea
The points $(h_c,v_c)$ and $(-h_c,-v_c)$, reached on $\Gamma$ when
$b=-\gamma$ (or $\gamma$) are the equivalent of ($\pm h_c, V=0$) in
the spin chain phase diagram. They,
again, cannot be approached from the flat phase by changing $v$ only,
since the line $h=h_c$ in the ($h,v$) plane is tangent to $\Gamma$. But,
from (\ref{def18}), near $b=-\gamma$
\bd
h_c - h\sim \half \biggl(\frac{I(k)}{\pi}\biggr)^{2}k^{\pr}(b+\gamma)^{2}
\ed
and since $f_{\mbox{\scriptsize step}}$ is linear in $b$ near
$b=-\gamma$ (unless u=0)
\bd
f_{\mbox{\scriptsize step}} \sim const (h_c -h)^{\half} + const (v-v_c)
\ed
at $(h_c,v_c)$. This shows that, like for the spin chain, the exponent
$\half$ dominates and signals that the points $(h_c,v_c)$ and $(-h_c,-v_c)$
are essentially different from the other points of $\Gamma$.

As to the sector $S^{z}=0$, we have to extend the numerical analysis
carried out for the spin chain. If $\Lambda_{0}(S^{z}=0,N,h,v_c)$
and $\Lambda_{1}(S^{z}=0,N,h,v_c)$ are the largest and next--to--largest
eigenvalues on the finite lattice in the sector under consideration,
we find that
\bd
\Delta \Lambda (S^{z}=0,h,v_c) =\lim_{N\ra\infty}\biggl[
\Lambda_{1}(S^{z}=0,N,h,v_c) -\Lambda_{0}(S^{z}=0,N,h,v_c)\biggr]
\ed
is positive for $h<h_c$ and vanishes when $h=h_c$. Furthermore
\bd
-\biggl[\ln \Lambda_{1}(S^{z}=0,N,h_c ,v_c) -\ln \Lambda_{0}
(S^{z}=0,N,h_c ,v_c)\biggr] \sim \frac{c^{\pr}}{N^{\half}}
\;\;\;\;\;\; N \gg 1
\ed
in perfect correspondence with the spin chain scaling of low--lying
excitations (see table $6$).
%
%
\section{The exponent $\frac{3}{2}$ of the free energy singularity}

As the field crosses the critical value of the $\Gamma$ line
(\ref{def18}), (\ref{def25})
the system enters a phase where horizontal and vertical polarizations
change continuously. This is an incommensurate phase belonging to the
universality class of the gaussian model \cite{Noh_Kim}. It is interesting
to determine the singularity of the free energy as ($h,v$) approach
$\Gamma$ from the incommensurate regime. It is widely believed
\cite{van_Beijeren, Nolden} that the free energy
singularity should be governed by an exponent $\frac{3}{2}$, but an
exact calculation has been done by Lieb and Wu when $h=0$ only
\cite{Lieb_Wu}, in which case
\bd
f\sim c(\gamma ,u)\biggl[ v-v_{c}(\gamma,u,b=0)\biggr]^{\frac{3}{2}}
\ed
Our calculation is an extension of Lieb's and Wu's method. We will
apply it first to the ground state energy of the spin chain, and
later extend if to the free energy of the statistical model.

Eqs. (\ref{def9}) and (\ref{def12})--(\ref{def13}) determine,
through the solution of (\ref{def11}),
$e_{0}$, $y$ and $h$ as functions of $A$ and $B$. We suppose that
such dependence is analytic and $e_{0}$, $y$ and $h$ can be expanded
in powers of $\delta A$, $\delta B$ as $A\ra A+\delta A$,
$B\ra B+\delta B$. Making explicit the dependence of $R(\alpha )$ on
$A$, $B$ by writing $R(\a ;A,B)$ we have, from (\ref{def12})
\ba
\partial_{A} y(A,B)&=&-\frac{2}{2\pi}\int_{A}^{B}d\a\partial_{A}
R(\a ;A,B) + \frac{2}{2\pi}R(A; A,B)\nonumber\\
\partial_{B} y(A,B)&=&-\frac{2}{2\pi}\int_{A}^{B}d\a\partial_{B}
R(\a ;A,B) - \frac{2}{2\pi}R(B; A,B)\nonumber\\
\delta y &=& \partial_{A} y(A,B)\delta A +
\partial_{B} y(A,B)\delta B + O(\delta A^2 ,\delta B^2,
\delta A\delta B)
\label{def45}
\ea
Likewise, the energy per site
\bd
e_{0} (A,B) = \cosh\gamma +\frac{1}{2\pi}\int_{A}^{B}d\a e(\a )
R(\a ;A,B) - Vy = \cosh\gamma - Vy + e_{0}^{(1)} (A,B)
\ed
yields the derivatives
\ba
\partial_{A}e_{0}^{(1)} (A,B) &=& \frac{1}{2\pi}\int_{A}^{B}d\a e(\a )
\partial_{A}R(\a ;A,B) -\frac{1}{2\pi}e(A)R(A;A,B)\nonumber\\
\partial_{B}e_{0}^{(1)} (A,B) &=& \frac{1}{2\pi}\int_{A}^{B}d\a e(\a )
\partial_{B}R(\a ;A,B) +\frac{1}{2\pi}e(B)R(B;A,B)\nonumber\\
\delta e_{0} (A,B) &=& \partial_{A} e_{0}^{(1)} (A,B)\delta A +
\partial_{B} e_{0}^{(1)} (A,B)\delta B + O(\delta A^2 ,\delta B^2,
\delta A\delta B)
\label{def46}
\ea
etc. Similar equations can be obtained for $h$, specializing (\ref{def9})
to the endpoints of the curve and taking the symmetric form
\bd
-4ih(A,B) = p^{0}(A)+p^{0}(B) - \frac{1}{2\pi}\int_{A}^{B}d\b
R(\b ;A,B)\biggr[ \Theta (A-\b )+\Theta (B-\b )\biggr]
\ed
hence
\ba
- 4 i \partial_{A} h(A,B) &=& -\frac{1}{2\pi}\int_{A}^{B} d\b \partial_{A}
R(\b ;A,B)
\biggr[ \Theta (A-\b )+\Theta (B-\b )\biggr]\nn \\
& & \;\;\;  + R(A;A,B)\biggr[1+\frac{1}{2 \pi}
\Theta(B-A)\biggl]\nn \\
- 4 i \partial_{B} h(A,B) &=& -\frac{1}{2\pi}\int_{A}^{B} d\b \partial_{B}
R(\b ;A,B)
\biggr[ \Theta (A-\b )+\Theta (B-\b )\biggr]\nn \\
& & \;\;\; + R(B;A,B)\biggr[1+\frac{1}{2 \pi}
\Theta(B-A)\biggl]\nn
\ea
etc.. Equations for the derivatives of $R(\alpha;A,B)$
are readily obtained from
\ba
\partial_{A} R(\alpha; A,B) + \frac{1}{2 \pi} \int_A^{B} d\b K(\alpha - \beta)
\partial_{A} R(\beta; A,B) &=& \frac{1}{2 \pi} K(\alpha - A) R(A; A,B)\nn\\
\partial_{B} R(\alpha; A,B) + \frac{1}{2 \pi} \int_A^{B} d\b K(\alpha - \beta)
\partial_{B} R(\beta; A,B) &=& - \frac{1}{2 \pi} K(\alpha - B) R(B; A,B)
\nn
\ea
etc..

We have carried out these expansions to the third order in 
$\delta A, \delta B$. In principle they can be used for any $A,B$
with $a = \pi, |b| \leq \gamma$, and the integrals computed by Fourier 
transform. However, as it is already evident from the first order terms,
the expansions simplify considerably when carried out around $A_0 = -\pi \pm
i \gamma, B_0 = \pi \pm i \gamma$ which are zeros of the $dn$ function
\bd
R(A_0;A_0,B_0) = R(B_0;A_0,B_0) =0
\ed
The details of the expansion are lengthy but straightforward, so only the final
form is of interest. Writing
\bd
\delta A = - \delta a + i \delta b \;\;\;\;\;\; 
\delta B = \delta a + i \delta b
\ed
and considering first the expansion around
$A_0 = -\pi - i \gamma, B_0 = \pi - i \gamma$, we have
\ba
\delta e_0 &=& 2 c_2 \bigl[(\delta a)^3 - 3 \delta a (\delta b)^2\bigr] 
-V \delta y+ \ldots\label{def47}\\
\d y &=& -\frac{c_1}{\pi} \d a \d b +\frac{c_3}{\pi} \d b (\d a)^2 + \ldots
\label{def48}
\\
\d h &=& \frac{c_1}{2}\bigl[ (\d a)^2 - (\d b)^2\bigr]
+ \frac{c_3}{3} (\d a)^3 + \ldots
\label{def49}
\ea
with
\ba
c_1 &=& k^{\prime} (\frac{I(k)}{\pi})^2 > 0\nn\\
\frac{c_2}{c_1} &=& \frac{\sinh \gamma}{3 \pi} \biggl[\frac{1}{4} + \sum_{n>0}
\frac{(-1)^{n} n \exp(-n \gamma)}{\cosh \gamma} \biggr] > 0 \nn\\
\frac{c_3}{c_1} &=& \frac{1}{\pi} \sum_{n} \frac{\exp(-|n| \gamma)}
{2 \cosh \gamma n}  > 0
\nn
\ea
Notice that $\d y = 0 = \d e_0$ when $\d a =0$, as it should, since by taking
$\d a = 0 $ and $\d b >0$ we reenter the `flat phase'. 
It is also important to check that if $\d a =0$ there is no way to increase $h$ by changing $b$,
as already discussed in section 2. An increase in $h$, when keeping $y$
fixed at $y=0$, can instead be achieved by $\d a <0, \d b=0$, 
which confirms the numerical findings presented in section 2. A variation
$\d a>0$ is ruled out a priori, because the periodicity of (\ref{def3}) in the
real direction implies that rapidities are contained in the strip $ - \pi
\leq Re(\alpha) \leq \pi$ and $a$ cannot exceed $\pi$. 

Another point to discuss
is the reliability of (\ref{def47})--(\ref{def49}) when $n > N/2 $,
that is $\d y <0$.
The Bethe-ansatz equations for the symmetric six vertex model are always 
discussed keeping $n \leq N/2 $, since the $Z_2$ symmetry of arrow reversal
guarantees that the spectrum is the same when $N/2 < n \leq N$. It is
not immediately clear what happens to (\ref{def3}) when $n > N/2$. As an 
example, consider the one dimensional sector $S^{z}=-N$ ($n =N$),
whose only eigenstate
is $| \downarrow \downarrow \downarrow \cdots \downarrow \rangle$. It is
not obvious that (\ref{def3}) should have only one solution when the number of 
unknowns is $N$. To be on the safe side we will trust
(\ref{def47})--(\ref{def49})
only for $n\leq \frac{N}{2}$ ($y \geq 0$). In this case $\d b \geq 0$ and
$\d a \leq 0$. To deal with the states at $n>\frac{N}{2}$ ($y <0$),
one must resort to (\ref{def58}) which implies
\be
e_0(\gamma, h_c +\d h, V, -y) = e_0 (\gamma, -h_c-\d h, -V, y)
\label{def50}
\ee
Hence it is necessary to consider also an expansion around $h= -h_c, y=0$. This
can be done evaluating (\ref{def45}) and the following equations at
the endpoints
$A_0^{\prime} = - \pi + i \g, B_{0}^{\prime} = \pi + i \gamma$. 
The result is that (\ref{def47})--(\ref{def49}) still hold, with
$\d b >0 \;\;(<0)$ if $\d y > 0 \;\;(<0$).
A final observation about (\ref{def47})--(\ref{def49})
is that it is legitimate to neglect higher order terms in
(\ref{def48}) and (\ref{def49}). The parameters $\d a$ and
$\d b$ are independent and there is no control over their relative magnitude,
but the second order term in (\ref{def49}) is dominant unless $\d a \simeq \pm 
\d b$, in which case the third order term is certainly larger than all possible
fourth order terms. Likewise, in (\ref{def48}), no term $\d a^n$ or $\d b^n$ is
allowed since we know that $\d y =0$ if $\d a =0$ or $\d b =0$. Consequently,
all higher order terms can certainly be neglected and one can further limit the
expansion to
\bd
\d y = - \frac{c_1}{\pi} \d a \d b
\ed
Suppose now that $h$ is kept fixed at $h_c$ and $V \neq 0$. Then, from 
(\ref{def49}), 
\bd
\d b = \mp \d a
\ed
where the upper (lower) sign holds for $\d y >0\;\; (<0)$. Consider first
$\d y >0$. Then
\ba
\d y &=& \frac{c_1}{\pi} \d a^2 \nn \\
\d e_0 (\d a) &=& - 4 c_2 \d a^3 - \frac{V c_1}{\pi} \d a^2
\nn
\ea
which has a minimum, when $V>0$, at
\bd
\d a_0 = - \frac{V c_1}{6 \pi c_2}
\ed
that yields
\bd
\d e_0(h=h_c,
V > 0) = - \frac{2}{c_2^2} \biggl(\frac{V c_1}{6\pi}\biggr)^3
\ed
Notice that no minimum occurs if $V<0$. Instead, if we consider $\d y <0$,
one has a minimum at 
\bd
\d a_0 =  \frac{V c_1}{6 \pi c_2}
\ed
that yields 
\bd
\d e_0(h=h_c,
V < 0) =  \frac{2}{c_2^2} \biggl(\frac{V c_1}{6\pi}\biggr)^3
\ed
when $V < 0$. Consequently, 
\ba
e_0(h=h_c, V) &=& e_0(h=h_c, V=0)-\frac{2}{c_2^2} 
\biggl(\frac{ c_1}{6\pi}\biggr)^3 |V|^3\nn \\
\d y &=& sgn(V) \frac{c_1}{\pi} \biggl(\frac{V c_1}{6 \pi c_2}\biggr)^2
\nn
\ea
is the ground state energy singularity as one approaches the point ($h_c,V=0$)
along the $V$ direction.

The case $V=0, \d h \neq 0$ is more involved. We want $\d h >0$, in order to
move into the incommensurate phase. From (\ref{def47})--(\ref{def49})
\ba
\d b &=& \pm \sqrt{f(\d a)} \;\;\;\;\;\;\;\;\;
 f(\d a) = \d a^2 + \frac{2}{3} \frac{c_3}
{c_1} \d a^3 - \frac{2}{c_1} \d h\label{def51} \\
\d y &=& - \frac{c_1}{\pi} \d a (\pm \sqrt{f(\d a)})
\label{def52}\\
\d e_0(\d a) &=& 2 c_2 \d a \biggl( -2 (\d a)^2 - 2 \frac{c_3}{c_1} 
(\d a)^3 +\frac{6}{c_1} \d h\biggr)
\label{def53}
\ea
where the sign in (\ref{def52}) depends on whether we want
$ \d y > 0$ or $ \d y < 0$. The variation $\d a$ must be negative and
contained in a range where $f(\d a)$ is non negative, so if
\bd
f(\d a_0) =0 \;\;\;\;\; \d a_0 = - \sqrt{\frac{2}{c_1}} (\d h)^{1/2} + O(\d h)
\ed
we consider
\bd
\d a \leq \d a_0
\ed
It is not difficult to see that there is a left neighborhood of $\d a_0$ (of
the order $\d h^{1/2}$) where
\begin{itemize}
\item[1.] $f(\d a )$ is positive
\item[2.] $\d y(\d a)$ is monotonic
\item[3.] $\d e_0(\d a)$ is decreasing 
\end{itemize}

Consequently, regardless of the sign in (\ref{def52}), $\d a = \d a_0$
is a local
minimum of $\d e_0(\d a)$, which, incidentally, corresponds to $\d y =0$.
Inserting $\d a_0$ in (\ref{def53})
\be
e_0(h_c+\d h, V=0) = e_0 (h_c, V=0)- 2 c_2 \biggl(\frac{2}{c_1} \d h
\biggr)^{3/2}
\label{def54}
\ee
Although it is not obvious from the previous proof, $\d y =0$ is actually a
stationary point for $\d e_0(\d y)$. In fact
\bd
\frac{\partial \d e_0}{\partial \d y} = \frac{\frac{\partial \d e_0}{\partial
\d a}\big\vert_{\d a_0}}{\frac{\partial \d y}{\partial \d a }
\big\vert_{\d a_0}} 
\ed
and
\bd
\frac{\partial \d y}{\partial \d a} = \pm\biggl( -\frac{c_1}{\pi}
\sqrt{f(\d a)}
- \frac{c_1}{2 \pi} \d a \frac{f^{\pr} (\d a)}{\sqrt{f(\d a)}}
\biggr)
\ed
becomes infinite at $\d a_0$.

The calculation of the free energy singularity is a simple extension of this
method. The variation of the ground state  energy is now replaced by 
(see (\ref{def23}))
\bd
- \d (F(u,\gamma,h,y)+v y) = - \d F - \d y(v_c+\d v)
\ed
The variation of $F$ is computed by means of an expansion analogous to
(\ref{def45})--(\ref{def46}). To keep things simple we
consider $d<\gamma - 2 u$, which
is certainly true for $u$ sufficiently small, so that $\Lambda_R$ dominates 
over $\Lambda_L$. The surprisingly simple result is that, like for the spin 
chain, the first nonzero contribution comes at the third order. The quantity
to minimize is 
\bd
2 c_2^{\pr} (\d a^3 - 3 \d a \d b^2) - \d y \d v
\ed
where
\bd
c_2^{\pr} = \frac{c_1}{6 \pi} \bigl[ \frac{1}{2} + \sum_{n>0} 
\frac{(-1)^n \cosh(\gamma(n-2 u))}{\cosh(\gamma n)} \bigr] >0
\ed
which looks exactly like (\ref{def47})--(\ref{def49})
provided $c_2 \rightarrow c_2^{\pr}$
and $V \rightarrow \d v$. The conclusions are therefore the same and the
free energy leading singularities approaching ($h_c,v_c$) from the
incommensurate phase are
\ba
f(u,\gamma, h_c+\d h, v_c) &=& f(u, \gamma, h_c, v_c)- 2 c_2^{\pr} \biggl(
\frac{2}{c_1} \d h\biggr)^{3/2} \nn \\
f(u, \gamma, h_c, v_c + \d v) &=& f(u, \gamma, h_c, v_c) -
\frac{2}{{c_{2}^{\pr}}^2} \biggl(\frac{c_1}{6\pi}\biggr)^3 |\d v|^3
\nn
\ea
%
%
\section{Discussion}

It is interesting to speculate about the nature of
the phase transition at ($h_c,v_c$) and compare it with what happens at
the other points of $\Gamma$.

We have always worked in the assumption that a complete set of
eigenstates exists for the transfer matrix. It is well known then that
the correlation function of $2$ vertical arrows can be analysed through
a spectral decomposition \cite{Johnson_Krinsky_McCoy}.
For the correlator of two vertical arrows
along the same column one has, on a $N\times M$ lattice
\be
\langle\a_{0,0}\;\a_{0,n}\rangle = \frac{Tr(\sig_{0}^{z}T^{n}\sig_{0}^{z}
T^{M-n})}{Tr T^{M}}\stackrel{M\ra\infty}{\longrightarrow}\sum_{k}\bigg\vert
\langle 0| \sig_{0}^{z} |k\rangle\bigg\vert^{2}\biggl(\frac
{\Lambda_k}{\Lambda_0}\biggr)^{n}
\label{def55}
\ee
and for the correlation function along the horizontal direction
\be
\langle\a_{0,0}\;\a_{n,0}\rangle = \frac{Tr(\sig_{0}^{z}\sig_{n}^{z}
T^{M})}{Tr T^{M}}\stackrel{M\ra\infty}{\longrightarrow}
\langle 0| \sig_{0}^{z}\sig_{n}^{z} |0\rangle
\label{def56}
\ee
where we have denoted with $\ket{0}$ the eigenstate of the largest
eigenvalue of the transfer matrix on a finite lattice of width $N$.
Here $\sum_k$ denotes the sum over a complete set of eigenvectors
of the transfer matrix. It is useful to consider first what happens
at $h=0$. Since, obviously
\bd
\biggl[ \sig_{n}^{z},\sum_{j=1}^{N}\sig_{j}^{z}\biggr]=0
\ed
and the ground state lies in the sector $S^{z}=0$, only this
sector contributes to (\ref{def55}) and (\ref{def56}). Furthermore, $v$ does
not change the spectrum within a sector of fixed $S^{z}$, and it does not
modify the eigenvectors which depend (as explicitly seen from the
Bethe Ansatz \cite{Lieb_Wu, Jayaprakash_Sinha})
on $\gamma$ and $h$ only. We conclude that
$v$ has no effect whatsoever on the correlators, which remain those of the
symmetric six--vertex model, until the level--crossing transition
takes place at $v=v(\gamma,u,b=0)$ (as given by (\ref{def25})). Here
the system moves into the gaussian phase, and the ground state even
for small $\delta v=v-v(\gamma,u,b=0)$ falls into a sector at $y\neq 0$
\cite{Lieb_Wu}, hence with $S^{z}$ of order $N$. The transition occurs
without divergence of the correlation length which jumps from the
(finite) value of the symmetric six--vertex to infinity.

It is our conjecture that this picture does not change when
$-h_c <h< h_c$, and the level crossing transition persists along all
$\Gamma$, up to the points ($h_c ,v_c$) or ($-h_c ,-v_c$). Here the
correlation length should diverge according to the following
argument. The fact that in the sector $S^{z}=0$, as found numerically
in section $3$,
\bd
\lim_{h\ra h_{c}^{-}}\Delta\ln\Lambda (S^{z}=0,h,v_c)=0
\ed
is, by itself, not sufficient to prove the divergence
of the correlation length $\xi_1$ in the vertical direction.
The fact that $\Lambda_{k}$ in (\ref{def55}) are generally complex forces one
to sum over a whole band of them because oscillations can affect the
behavior of $\xi_1$ \cite{Johnson_Krinsky_McCoy}.
This in not doable until it is clarified
what takes the place of strings in the sector under consideration.
Nevertheless, one can look at the horizontal correlation function
in (\ref{def56}). As shown in section $4$, it is possible to enter the
gaussian phase at ($h_c,v_c$) keeping the ground state at $y=0$.
From (\ref{def10}) $y=0$ does not necessarily imply that $S^{z}=0$
($n=\frac{N}{2}$), since $S^{z}$ can be nonzero but remain finite
in the limit $N\ra\infty$ and $y$ would still be zero. Still it is
tempting to conjecture that the ground state remains at $S^{z}=0$,
and this is confirmed by preliminary numerical results on the spin
chain. Hence no level crossing occurs here, because the Perron--Frobenius
theorem prevents if from happening in a sector of fixed $S^{z}$,
and the expectation value in (\ref{def56})
is taken with the same Bethe--Ansatz eigenvector in the $2$ phases.
Since in the gaussian phase the horizontal correlation lenght $\xi_2$
is infinite, $\xi_{2}(h)$ must diverge as $h\ra\pm h_c$.

Although these conclusions are rather speculative, they seem to warrant
a further investigation of the phase transition at issue. It should be
recalled that, if indeed ($h_c,v_c$) is critical in the sense of
diverging correlation length, it cannot be classified in terms of
conformal field theory as shown by the anomalous scaling discussed in
section $3$.\\

\vspace{1.0cm}

\noindent
{\Large \bf Acknowledgements}

We wish to thank Prof. C. Destri, Prof. G. von Gehlen
and Prof. P. Pearce for useful discussions. Special thanks go to Prof.
V. Rittenberg for many discussions and constant encouragement. One of us
(GA) has been supported by the EC under the program 'Human Capital and
Mobility.'
\section*{Appendix A}
We discuss here the symmetries of $\cal H$ and of the transfer matrix.
Under the action of the (unitary)  charge conjugation operator
\bd 
C = \prod_{k=1}^{N}\sigma_{k}^{x} \;\;\;\;\; C=C^{\dagger}=C^{-1}
\ed
$\cal H$ transforms as 
\be 
C {\cal H} (\gamma, h, V) C = {\cal H} (\gamma,-h, -V)
\label{def57}
\ee
Since
\bd 
C\; (\sum_{j=1}^{N}\sigma_{j}^{z})\; C = - \sum_{j=1}^{N}\sigma_{j}^z
\ed
the spectra of the sectors at fixed $S^{z}$ are related by
\be
{\cal S} (\gamma,h,V,S^{z}) = {\cal S}(\gamma, -h, -V, -S^{z})
\label{def58}
\ee
The same symmetry operation can be applied to the transfer matrix, using 
the matrix form of $C$
\bd
\langle {\underline \alpha}, C {\underline \alpha^{\prime}}\rangle = 
\prod_{k=1}^{N} \delta_{\alpha_k, - \alpha_k^{\prime}} 
\ed
From (\ref{def1}) and the definition of the Botzmann weights it is elementary
to see that 
\be
C \; T(\gamma, u, h,v) \; C \bigg\vert_{\underline \alpha, \underline 
\alpha^{\prime}}
= T(\gamma,u,-h,-v) \bigg\vert_{\underline \alpha, \underline \alpha^{\prime}}
\label{def59}
\ee
which implies the relation between spectra
\be
{\cal S}_{TM} \{h,v,S^{z}\} = {\cal S}_{TM} \{ -h, -v, -S^{z} \}
\label{def60}
\ee
This symmetry manifests itself in the fact that the partition function 
\bd
Z_{PF}(h,v) = Z_{PF}(-h, -v)
\ed
As far as $\cal H$ is concerned, another symmetry is in effect, implemented by
the space inversion operator (not to be confused with the momentum)
\bd
P |\alpha_1, \alpha_2,\ldots,\alpha_{N-1}, \alpha_{N}\rangle =
|\alpha_{N}, \alpha_{N-1},\ldots,\alpha_2,\alpha_1 \rangle\;\;\;P^2 = 1
\ed
\bd
P\;{\cal H}(\gamma, h, V)\;P = {\cal H}(\gamma,-h,V)
\ed
Hence
\bd
C\; P \;{\cal H} (\gamma, h, V=0) \;P \;C = {\cal H} (\gamma, h, V=0)
\ed
So at $V=0$ the spin chain recovers the $Z_2$ symmetry under spin reversal
and the spectra at $S^{z}$ and $-S^{z}$ are identical. It is noteworthy
that the same is not true for the transfer matrix.

\section*{Appendix B}
Our Hamiltonian, at $h=0$, $V=0$, can be written, neglecting a constant 
additive term
\be
{\cal H} = -\frac{1}{2}\sum_{j=1}^{N}\biggl[ \Delta 
\sigma^z_j \sigma^z_{j+1}
              +  \sigma^x_j \sigma^x_{j+1} + 
\sigma^y_j \sigma^y_{j+1} \biggr]\;\;\;\;\;\mbox{with}\;\;\;\;\;\Delta < 1
\label{def61}
\ee
and it is mapped onto the Hamiltonian (see, {\it e.g.} \cite{Gaudin})
\be
{\cal H}_{G} = \frac{1}{2}\sum_{j=1}^{N}\biggl[ \Delta 
\sigma^z_j \sigma^z_{j+1}
              +  \sigma^x_j \sigma^x_{j+1} + 
\sigma^y_j \sigma^y_{j+1} \biggr]\;\;\;\;\;\mbox{with}\;\;\;\;\;\Delta>1
\label{def62}
\ee
through a unitary transformation
\bd
{\cal H}_{G} = U\; {\cal H} \;U^{-1} \;\;\;\;\;\;
U = \prod_{j=1}^{N/2} \sigma^{z}_{2j-1}
\ed
Consider the shift operator $S$ of (\ref{def6}). Let $| \Psi_n, P \rangle$
be an eigenstate of $\cal H$ with momentum $P$ and $S^{z}=\frac{N}{2} -n$
\ba
\half\sum_{j=1}^{N}\;\; \sigma^z_j\;\; | \Psi_n, P \rangle &=&(\frac{N}{2}-n)
\;\; | \Psi_n, P \rangle\nonumber\\
S\;\;| \Psi_n, P \rangle &=& \exp(- i P) \;\; | \Psi_n, P \rangle
\nn
\ea
Define $U_0$ as
\bd
U_0 \;\; | \Psi_n, P \rangle := \prod_{j=1}^{N} \sigma_j^{z} \;\; | \Psi_n, 
P \rangle
= (-1)^n \;\;| \Psi_n, P \rangle
\ed
Then, an eigenstate of ${\cal H}_G$ with the same energy is $U \;\;| 
\Psi_n, P \rangle$. Since
\ba
S\;\; U \;\;| \Psi_n, P \rangle & = & \exp(-i P) \;\;S \;\;U \;\;S^{-1} \;\;
| \Psi_n, P \rangle\nonumber\\ 
&=&\exp(-i P) \;\;\prod_{j=1}^{N/2}\;\;\sigma^z_{2 j}\;\;| \Psi_n, P \rangle
\nn \\
\exp(-i P) \;\; U \;\;U_0 \;\;| \Psi_n, P \rangle  & = & \exp(-i(P+n \pi)) \;\;
U \;\;| \Psi_n, P \rangle\nonumber\\
&=&\exp(-i \overline{P}) \;\;U \;\;
| \Psi_n\rangle
\nn
\ea
$U \;\; | \Psi_n, P \rangle$ has momentum $\overline{P}=P + n \pi$. 
Hence the ground state of (\ref{def62})
has momentum $\overline{P_0}= n_0\;\pi= \frac{N}{2} \pi$,
while $P_0 =0$. As to the excitations
\bd
\Delta {\overline P} = \Delta P + (n-n_0)\pi = \Delta P - \frac{N_{h}\pi}{2}
\ed
Each hole carries an additional momentum $- \pi/2$ which implies
\bd
\Delta E = m_0 \sqrt{1-k^2 \sin^2 (\Delta P)} \;\;\;\;\rightarrow  \;\;\;\Delta E = m_0
\sqrt{1 - k^2 \cos^2 (\Delta P)}
\ed
%
%

\pagebreak
\noindent
{\bf List of Tables}
\noindent
\begin{enumerate}
\item[1.]
 Position of the endpoints $\pm a + i b$ of the curve $C$ for $h<h_c$;
$\exp(2h) = 3$
\item[2.]
Position of the endpoints $\pm a + i b$ of the curve $C$ for $h>h_c$;
$\exp(2h) = 18 $
\item[3.]
Energy gap of the spin chain for the first excited state
for different values of $h$ compared to the analytical result (3.15)
\item[4.]
Mass gap of the spin chain for different values of $h \leq h_c$
\item[5.]
Exponent of  $N$ for the finite-size corrections of the energy
gap in the sector $S^z = 0$ on the critical line $h=h_c$
\item[6.]
Exponent for the scaling of the free energy gap in the sector
$S^z = 0$ 
\end{enumerate}
\pagebreak
%
%
%
\vskip  0.8cm
\noindent
\ba
\begin{array}{||l||l|l||}\hline
\multicolumn{1}{||c||}{\mbox{\bf\sf Lattice size}}
&\multicolumn{1}{|c|}{a}
&\multicolumn{1}{|c||}{b}\\
\hline\hline
%
%
\mbox{\scriptsize $8$}&
\mbox{\scriptsize $ 2.226901585$}&
\mbox{\scriptsize $-1.186380535$}
\\ \hline
%
%
\mbox{\scriptsize $12$}&
\mbox{\scriptsize $2.522418777$}&
\mbox{\scriptsize $-1.215745628$}
\\ \hline
%
%
\mbox{\scriptsize $16$}&
\mbox{\scriptsize $2.674453171$}&
\mbox{\scriptsize $-1.227403536$}
\\ \hline
%
%
\mbox{\scriptsize $20$}&
\mbox{\scriptsize $2.766810718$}&
\mbox{\scriptsize $-1.233071106$}
\\ \hline
%
%
\mbox{\scriptsize $24$}&
\mbox{\scriptsize $2.828778340$}&
\mbox{\scriptsize $-1.236225697$}
\\ \hline
%
%
\mbox{\scriptsize $28$}&
\mbox{\scriptsize $2.873206253$}&
\mbox{\scriptsize $-1.238154458$}
\\ \hline
%
%
\mbox{\scriptsize $32$}&
\mbox{\scriptsize $2.906605731$}&
\mbox{\scriptsize $-1.239417236$}
\\ \hline
%
%
\mbox{\scriptsize $36$}&
\mbox{\scriptsize $2.932624180$}&
\mbox{\scriptsize $-1.240288031$}
\\ \hline
%
%
\mbox{\scriptsize $40$}&
\mbox{\scriptsize $2.953462078$}&
\mbox{\scriptsize $-1.240913442$}
\\ \hline\hline
%
%
\mbox{\bf\sf Extrap. $\infty$}&
\mbox{\scriptsize $3.141592659 (9)$}&
\mbox{\scriptsize $-1.243603723 (1)$}
\\ \hline
%
%
&
\mbox{\scriptsize $\pi = 3.141592653$}&
\mbox{\scriptsize $-\gamma = -3.737102242$}
\\ \hline
\end{array}
\nn
\ea
\vskip 0.5cm
\noindent
Table 1: Position of the endpoints $\pm a + i b$ of the curve $C$ for $h<h_c$;\\
$\exp(2h) = 3$
%
%
%
%
%
\vskip  0.8cm
\noindent
\ba
\begin{array}{||l||l|l||}\hline
\multicolumn{1}{||c||}{\mbox{\bf\sf Lattice size}}
&\multicolumn{1}{|c|}{a}
&\multicolumn{1}{|c||}{b}\\
\hline\hline
%
%
\mbox{\scriptsize $8$}&
\mbox{\scriptsize $1.345603261 $}&
\mbox{\scriptsize $-3.0497549071$}
\\ \hline
%
%
\mbox{\scriptsize $12$}&
\mbox{\scriptsize $1.479730321$}&
\mbox{\scriptsize $-3.2366686985$}
\\ \hline
%
%
\mbox{\scriptsize $16$}&
\mbox{\scriptsize $1.535932819$}&
\mbox{\scriptsize $-3.3472122872$}
\\ \hline
%
%
\mbox{\scriptsize $20$}&
\mbox{\scriptsize $1.564274277$}&
\mbox{\scriptsize $-3.4190538006$}
\\ \hline
%
%
\mbox{\scriptsize $24$}&
\mbox{\scriptsize $1.580405494$}&
\mbox{\scriptsize $-3.4690826450$}
\\ \hline
%
%
\mbox{\scriptsize $28$}&
\mbox{\scriptsize $1.590409443$}&
\mbox{\scriptsize $-3.5057653322$}
\\ \hline
%
%
\mbox{\scriptsize $32$}&
\mbox{\scriptsize $1.597021761$}&
\mbox{\scriptsize $-3.5337453681$}
\\ \hline
%
%
\mbox{\scriptsize $36$}&
\mbox{\scriptsize $1.601612009$}&
\mbox{\scriptsize $-3.5557588049$}
\\ \hline
%
%
\mbox{\scriptsize $40$}&
\mbox{\scriptsize $1.604924714$}&
\mbox{\scriptsize $-3.5735136386$}
\\ \hline\hline
%
%
\mbox{\bf\sf Extrap. $\infty$}&
\mbox{\scriptsize $1.6193362 (3)$}&
\mbox{\scriptsize $-3.737102232  (7)$}
\\ \hline
%
%
&
\mbox{\scriptsize $\pi = 3.141592653$}&
\mbox{\scriptsize $-\gamma = -3.737102242$}
\\ \hline
\end{array}
\nn
\ea
\vskip 0.5cm
\noindent
Table 2: Position of the endpoints $\pm a + i b$ of the curve $C$ for $h>h_c$;\\
$\exp(2h) = 18$ 
%
%
%
%
%
\vskip  0.8cm
\noindent
\ba
\begin{array}{||l||l|l||}\hline
\multicolumn{1}{||c||}{\mbox{\bf\sf Lattice size}}
&\multicolumn{1}{|c|}{\mbox{\bf\sf $e^{2h}=1.0$}}
&\multicolumn{1}{|c||}{\mbox{\bf\sf $e^{2h}=9.0$}}\\
\hline\hline
%
%
\mbox{\scriptsize $8$}&
\mbox{\scriptsize $38.8549946261240$}&
\mbox{\scriptsize $27.3592680025296$}
\\ \hline
%
%
\mbox{\scriptsize $12$}&
\mbox{\scriptsize $38.4729554492734$}&
\mbox{\scriptsize $23.5353424072427$}
\\ \hline
%
%
\mbox{\scriptsize $16$}&
\mbox{\scriptsize $38.3095545699746$}&
\mbox{\scriptsize $21.4216001709441$}
\\ \hline
%
%
\mbox{\scriptsize $20$}&
\mbox{\scriptsize $38.2251709688714$}&
\mbox{\scriptsize $20.0937869680560$}
\\ \hline
%
%
\mbox{\scriptsize $24$}&
\mbox{\scriptsize $38.1760618666305$}&
\mbox{\scriptsize $19.1948069770300$}
\\ \hline
%
%
\mbox{\scriptsize $28$}&
\mbox{\scriptsize $38.1450125441243$}&
\mbox{\scriptsize $18.5549980007490$}
\\ \hline
%
%
\mbox{\scriptsize $32$}&
\mbox{\scriptsize $38.1241500821837$}&
\mbox{\scriptsize $18.0828136699350$}
\\ \hline
%
%
\mbox{\scriptsize $36$}&
\mbox{\scriptsize $38.1094640015976$}&
\mbox{\scriptsize $17.7244227601235$}
\\ \hline
%
%
\mbox{\scriptsize $40$}&
\mbox{\scriptsize $38.0987382252780$}&
\mbox{\scriptsize $17.4461598430727$}
\\ \hline\hline
%
%
\mbox{\bf\sf Extrap. $\infty$}&
\mbox{\scriptsize $38.04991359 (1)$}&
\mbox{\scriptsize $15.8884 (8)$}
\\ \hline
%
%

\mbox{\bf\sf Exact eq.}&
\mbox{\scriptsize $38.04991361$}&
\mbox{\scriptsize $15.8887254$} 
\\ \hline
\end{array}
\nn
\ea
\vskip 0.5cm
\noindent
Table 3: Energy gap of the spin chain for the first excited state
for different values of $h$ compared to the analytical result (3.15)
%
%
%
%
%
\vskip  0.8cm
\noindent
\ba
\begin{array}{||l||l|l|l|l||}
\hline
\multicolumn{1}{||c||}{\mbox{\bf\sf Lattice size}}&
\multicolumn{1}{|c|}{e^{2h} = 9.0}&
\multicolumn{1}{|c|}{e^{2h} = 9.5}&
\multicolumn{1}{|c|}{e^{2h} = 10.0}&
\multicolumn{1}{|c||}{e^{2h_c} = 10.51787}\\
\hline\hline
%
%
\mbox{\scriptsize $8$}&
\mbox{\scriptsize $40.28463586426577$}&
\mbox{\scriptsize $40.60665243638542$}&
\mbox{\scriptsize $41.00560893816288$}&
\mbox{\scriptsize $41.50186071099847$}
\\ \hline
%
%
\mbox{\scriptsize $12$}&
\mbox{\scriptsize $32.50806556257382$}&
\mbox{\scriptsize $32.33242547913501$}&
\mbox{\scriptsize $32.27369006467646$}&
\mbox{\scriptsize $32.34509918439876$}
\\ \hline
%
%
\mbox{\scriptsize $16$}&
\mbox{\scriptsize $28.03119022066448$}&
\mbox{\scriptsize $27.47527007849106$}&
\mbox{\scriptsize $27.06291429820305$}&
\mbox{\scriptsize $26.80804074773090$}
\\ \hline
%
%
\mbox{\scriptsize $20$}&
\mbox{\scriptsize $25.23278497081201$}&
\mbox{\scriptsize $24.37422080701744$}&
\mbox{\scriptsize $23.67516261790179$}&
\mbox{\scriptsize $23.15771126310622$}
\\ \hline
%
%
\mbox{\scriptsize $24$}&
\mbox{\scriptsize $23.35422673956813$}&
\mbox{\scriptsize $22.24638140499817$}&
\mbox{\scriptsize $21.30438602193786$}&
\mbox{\scriptsize $20.56522041772249$}
\\ \hline
%
%
\mbox{\scriptsize $28$}&
\mbox{\scriptsize $22.02474072486461$}&
\mbox{\scriptsize $20.70732422258583$}&
\mbox{\scriptsize $19.55338243467817$}&
\mbox{\scriptsize $18.62033939778466$}
\\ \hline
%
%
\mbox{\scriptsize $32$}&
\mbox{\scriptsize $21.04552971555199$}&
\mbox{\scriptsize $19.54981005828927$}&
\mbox{\scriptsize $18.20747202515815$}&
\mbox{\scriptsize $17.10061549106331$}
\\ \hline
%
%
%
\mbox{\scriptsize $36$}&
\mbox{\scriptsize $20.30141022559842$}&
\mbox{\scriptsize $18.65287555243093$}&
\mbox{\scriptsize $17.14104807658379$}&
\mbox{\scriptsize $15.87553491911761$}
\\ \hline
%
%
\mbox{\scriptsize $40$}&
\mbox{\scriptsize $19.72145517781357$}&
\mbox{\scriptsize $17.94130164070793$}&
\mbox{\scriptsize $16.27579201027032$}&
\mbox{\scriptsize $14.86352304583235$}

\\ \hline\hline
%
%
\mbox{\bf\sf Extrap. $\infty$}&
\mbox{\scriptsize $15.946 (3)$}&
\mbox{\scriptsize $13.054 (9)$}&
\mbox{\scriptsize $9.37499 (4)$}&
\mbox{\scriptsize $ -1.3(3)\cdot 10^{-5} $}
\\ \hline
%
%
\end{array}\nonumber
\ea
\vskip 0.5cm
\noindent
Table 4: Mass gap of the spin chain for different values of $h \leq h_c$
\pagebreak
\newpage
%
%
%
\vskip  0.8cm
\noindent
\ba
\begin{array}{||l||l||}
\hline
\multicolumn{1}{||c||}{\mbox{\bf\sf Lattice size}}
&\multicolumn{1}{|c||}{\mbox{\bf\sf Exponent}}\\
\hline\hline
%
%
\mbox{\scriptsize $8$}&
\mbox{\scriptsize $-0.6147896371667922$}
\\ \hline
%
%
\mbox{\scriptsize $12$}&
\mbox{\scriptsize $-0.6526671750014708$}
\\ \hline
%
%
\mbox{\scriptsize $16$}&
\mbox{\scriptsize $-0.6559635919155932$}
\\ \hline
%
%
\mbox{\scriptsize $20$}&
\mbox{\scriptsize $-0.6511929107450443$}
\\ \hline
%
%
\mbox{\scriptsize $24$}&
\mbox{\scriptsize $-0.6444786397317260$}
\\ \hline
%
%
\mbox{\scriptsize $28$}&
\mbox{\scriptsize $-0.6376031966136601$}
\\ \hline
%
%
\mbox{\scriptsize $32$}&
\mbox{\scriptsize $-0.6311198906745727$}
\\ \hline
%
%
\mbox{\scriptsize $36$}&
\mbox{\scriptsize $-0.6251786621408798$}
\\ \hline
%
%
\mbox{\scriptsize $40$}&
\mbox{\scriptsize $-0.6197881465618827$}
\\ \hline\hline
%
%
\mbox{\bf\sf Extrap. $\infty$}&
\mbox{\scriptsize $-0.5031 (1)$}
\\ \hline
%
%
\mbox{\bf\sf expected value}&
\mbox{\scriptsize $-0.5$}
\\ \hline
\end{array}\nonumber
\ea
\vskip 0.5cm
\noindent
Table 5: Exponent of  $N$ for the finite-size corrections of the energy
gap in the sector $S^z = 0$ on the critical line $h=h_c$
\pagebreak
\newpage
%
%
%
%
\vskip  0.8cm
\noindent
\ba
\begin{array}{||l||l||}\hline
\multicolumn{1}{||c||}{\mbox{\bf\sf Lattice size}}
&\multicolumn{1}{|c||}{\mbox{\bf\sf Exponent}}\\
\hline\hline
%
%
\mbox{\scriptsize $8$}&
\mbox{\scriptsize $-0.6696077334198590$}
\\ \hline
%
%
\mbox{\scriptsize $12$}&
\mbox{\scriptsize $-0.6362863326795953$}
\\ \hline
%
%
\mbox{\scriptsize $16$}&
\mbox{\scriptsize $-0.6165741467010993$}
\\ \hline
%
%
\mbox{\scriptsize $20$}&
\mbox{\scriptsize $-0.603244670912252$}
\\ \hline
%
%
\mbox{\scriptsize $24$}&
\mbox{\scriptsize $-0.5934852417528157$}
\\ \hline
%
%
\mbox{\scriptsize $28$}&
\mbox{\scriptsize $-0.5859536198128706$}
\\ \hline
%
%
\mbox{\scriptsize $32$}&
\mbox{\scriptsize $-0.5799199948938484$}
\\ \hline
%
%
\mbox{\scriptsize $36$}&
\mbox{\scriptsize $-0.5749497141578801$}
\\ \hline
%
%
\mbox{\scriptsize $40$}&
\mbox{\scriptsize $-0.5707659045338731$}
\\ \hline\hline
%
%
\mbox{\bf\sf Extrap. $\infty$}&
\mbox{\scriptsize $-0.499998(4)$}
\\ \hline
%
%
\end{array}
\nn
\ea
\vskip 0.5cm
\noindent
Table 6: Exponent for the scaling of the free energy gap in the sector
$S^z = 0$ 
%
\newpage
%
%
%
%
\listoffigures
\newpage
\begin{figure}[h]
\setlength{\unitlength}{5mm}
\begin{picture}(54,24)
\put(0,0){\framebox(32,24)}
\put(3,1){\vector(0,1){1}}
\put(3,3){\vector(0,-1){1}}
\put(3,2){\vector(1,0){1}}
\put(3,2){\vector(-1,0){1}}
\put(3,6){\vector(0,1){1}}
\put(3,6){\vector(0,-1){1}}
\put(4,6){\vector(-1,0){1}}
\put(2,6){\vector(1,0){1}}
\put(3,10){\vector(0,-1){1}}
\put(3,11){\vector(0,-1){1}}
\put(3,10){\vector(1,0){1}}
\put(2,10){\vector(1,0){1}}
\put(3,13){\vector(0,1){1}}
\put(3,14){\vector(0,1){1}}
\put(3,14){\vector(-1,0){1}}
\put(4,14){\vector(-1,0){1}}
\put(3,18){\vector(0,-1){1}}
\put(3,19){\vector(0,-1){1}}
\put(3,18){\vector(-1,0){1}}
\put(4,18){\vector(-1,0){1}}
\put(3,21){\vector(0,1){1}}
\put(3,22){\vector(0,1){1}}
\put(2,22){\vector(1,0){1}}
\put(3,22){\vector(1,0){1}}
\put (6,1.8) {=}
\put (6,5.8) {=}
\put (6,9.8) {=}
\put (6,13.8) {=}
\put (6,17.8) {=}
\put (6,21.8) {=}
\put (8,1.8) {1}
\put (8,5.8) {1}
\put (8,9.8) {$e^{h-v}\;\frac{\sinh u}{\sinh \gamma} $}
\put (8,13.8) {$e^{-h+v}\;\frac{\sinh u}{\sinh \gamma}$}
\put (8,17.8) {$e^{-h-v}\;\frac{\sinh(\gamma-u)}{\sinh \gamma}$}
\put (8,21.8) {$e^{h+v}\;\frac{\sinh(\gamma-u)}{\sinh \gamma}$ }
\put (13,1.8) {=}
\put (13,5.8) {=}
\put (13,9.8) {=}
\put (13,13.8) {=}
\put (13,17.8) {=}
\put (13,21.8) {=}
\put (15,1.8) {1}
\put (15,5.8) {1}
\put (15,9.8)  {$ e^{\beta (-\delta/2 - \epsilon) + h - v} $}
\put (15,13.8) {$ e^{\beta (-\delta/2 - \epsilon) - h + v} $}
\put (15,17.8) {$ e^{\beta (\delta/2 - \epsilon) - h - v}$}
\put (15,21.8) {$ e^{\beta (\delta/2 - \epsilon) + h + v}$ }
\put (21,1.8) {=}
\put (21,5.8) {=}
\put (21,9.8) {=}
\put (21,13.8) {=}
\put (21,17.8) {=}
\put (21,21.8) {=}
\put (23,1.8) {$R^{21}_{12} (u)$}
\put (23,5.8) {$R^{12}_{21} (u)$}
\put (23,9.8) {$R^{11}_{22} (u)$}
\put (23,13.8) {$R^{22}_{11} (u)$}
\put (23,17.8) {$R^{22}_{22} (u)$}
\put (23,21.8) {$R^{11}_{11} (u)$}
\put (1.5,1.8) {{\scriptsize 2}}
\put (2.9,0.5) {{\scriptsize 1}}
\put (2.9,3.3) {{\scriptsize 2}}
\put (4.3,1.8) {{\scriptsize 1}}
\put (1.5,5.8) {{\scriptsize 1}}
\put (2.9,4.5) {{\scriptsize 2}}
\put (2.9,7.3) {{\scriptsize 1}}
\put (4.3,5.8) {{\scriptsize 2}}
\put (1.5,9.8) {{\scriptsize 1}}
\put (2.9,8.5) {{\scriptsize 2}}
\put (2.9,11.3) {{\scriptsize 2}}
\put (4.3,9.8) {{\scriptsize 1}}
\put (1.5,13.8) {{\scriptsize 2}}
\put (2.9,12.5) {{\scriptsize 1}}
\put (2.9,15.3) {{\scriptsize 1}}
\put (4.3,13.8) {{\scriptsize 2}}
\put (1.5,17.8) {{\scriptsize 2}}
\put (2.9,16.5) {{\scriptsize 2}}
\put (2.9,19.3) {{\scriptsize 2}}
\put (4.3,17.8) {{\scriptsize 2}}
\put (1.5,21.8) {{\scriptsize 1}}
\put (2.9,20.5) {{\scriptsize 1}}
\put (2.9,23.3) {{\scriptsize 1}}
\put (4.3,21.8) {{\scriptsize 1}}
\end{picture}
\caption{Boltzmann weights in the notation with spectral parameter $u$ 
compared to that of ref. $[9]$. The physical region is $0 < u < \gamma$} 
\label{picture}
\end{figure}
\thispagestyle{empty}
\newpage

{
\begin{figure}[h]
\def\sgr{\scriptstyle}
\setlength{\unitlength}{1mm}
\def\setl{ \setlength\epsfxsize{10.0cm}}
\begin{picture}(150,150)
\put(85,-3){\makebox{$Re(\alpha)$}}
\put(5,55){\makebox{$Im(\alpha)$}}
\put(10,100){
        \makebox{
                \setl
                \epsfbox{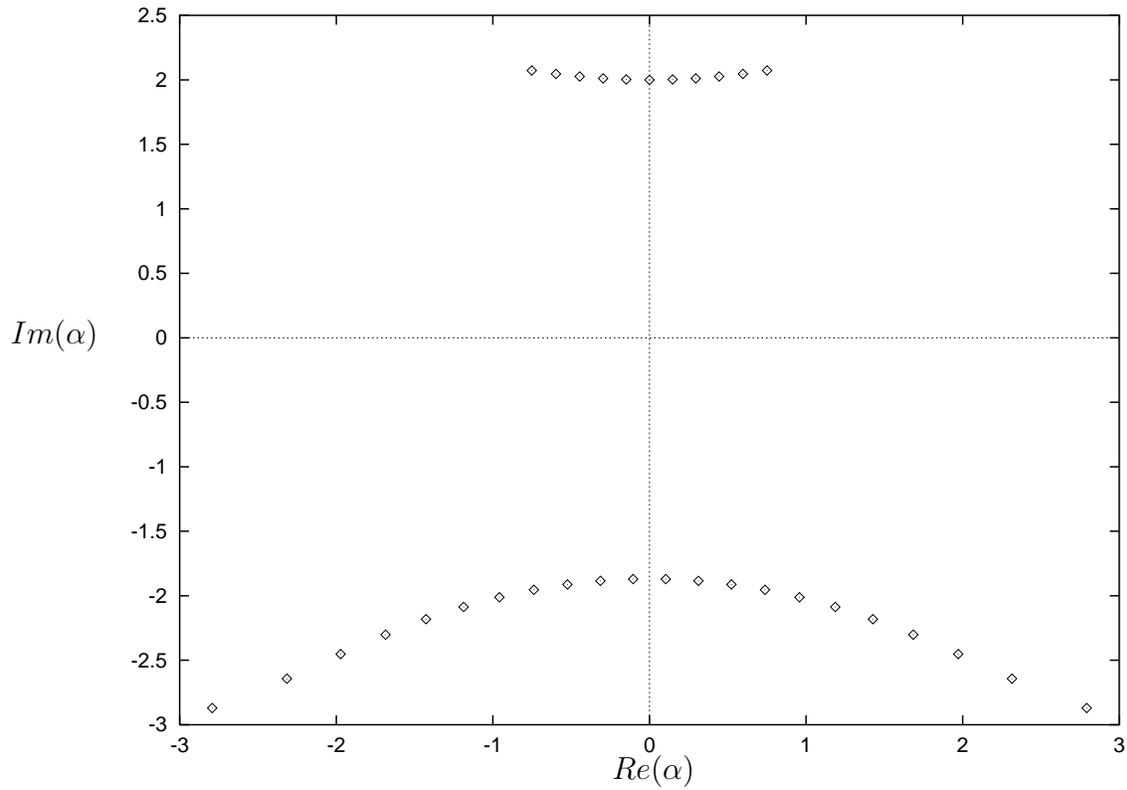}}
        }
\end{picture}
\caption{Distribution of the rapidities $\alpha_k$ for $N=44$. The lower
curve corresponds to the ground state in the sector with $n=\frac{N}{2}$
and $\exp (2 h) = 9$, the upper curve to the ground state of the sector
with $n=\frac{N}{4}$ and  $\exp (2 h) = \frac{1}{9}$}

\end{figure}
}

\pagebreak

\newpage

{
\begin{figure}[h]
\def\sgr{\scriptstyle}
\setlength{\unitlength}{1mm}
\def\setl{ \setlength\epsfxsize{10.0cm}}
\begin{picture}(150,150)
\put(10,100){
        \makebox{
                \setl
                \epsfbox{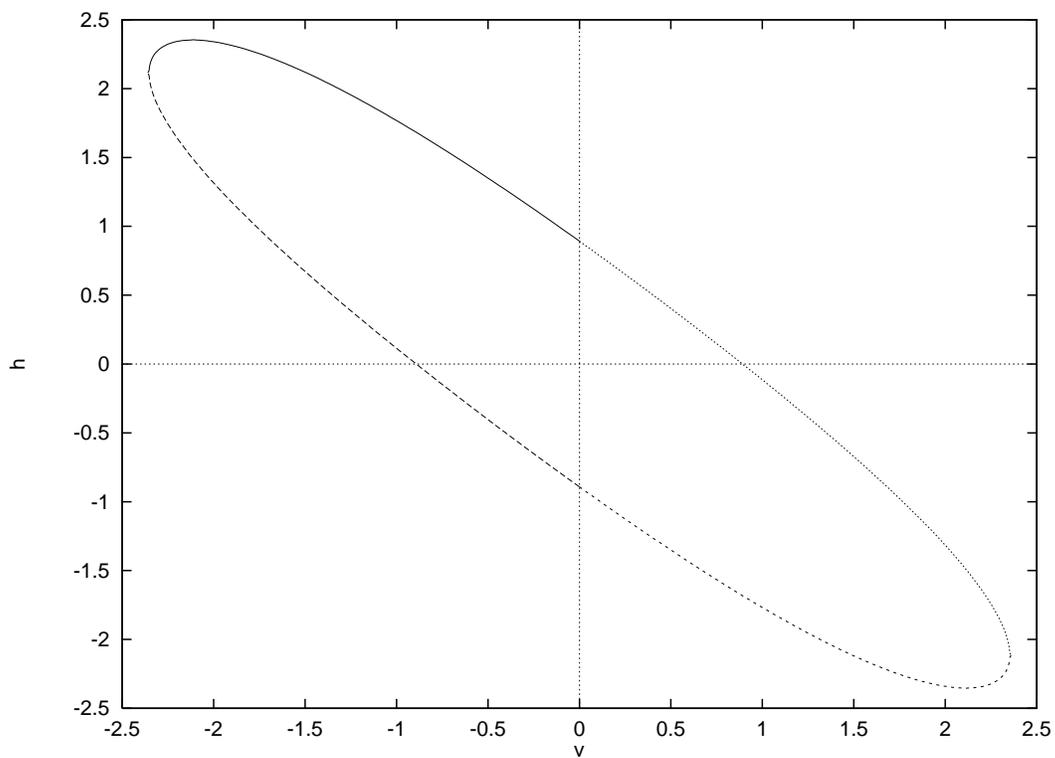}}
        }
\end{picture}
\caption{Curve $\Gamma$ in the $h-v-$plane for $u = \frac{1}{2}$
and $\cosh(\gamma) = 21$}
\end{figure}
}

\pagebreak
\newpage

\begin{figure}[h]
\def\sgr{\scriptstyle}
\setlength{\unitlength}{1mm}
\def\setl{ \setlength\epsfxsize{14.0cm}}
\begin{picture}(100,50)
\put(0,-90){
        \makebox{
                \setl
                \epsfbox{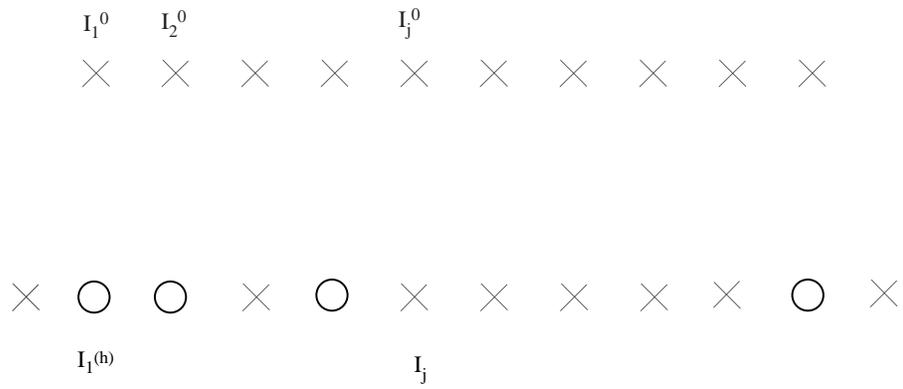}}
        }
\end{picture}
\caption{Distribution of $I_k$'s for the ground state and an excited 
state with $n=n_0-2$ rapidities}
\end{figure}
\end{document}